\documentclass[12pt]{article}
\usepackage{epsfig}

\begin{document}

\begin{titlepage}

\begin{flushright}
CLNS~04/1885\\
{\tt hep-ph/0408179}\\[0.2cm]
August 16, 2004
\end{flushright}

\vspace{0.7cm}
\begin{center}
\Large\bf\boldmath 
Renormalization-Group Improved Calculation of the\\
$B\to X_s\gamma$ Branching Ratio\unboldmath
\end{center}

\vspace{0.8cm}
\begin{center}
{\sc Matthias Neubert}\\
\vspace{0.7cm}
{\sl Institute for High-Energy Phenomenology\\
Newman Laboratory for Elementary-Particle Physics, Cornell University\\
Ithaca, NY 14853, U.S.A.}
\end{center}

\vspace{1.0cm}
\begin{abstract}
\vspace{0.2cm}\noindent
Using results on soft-collinear factorization for inclusive $B$-meson decay 
distributions, a systematic study of the partial $B\to X_s\gamma$ decay rate 
with a cut $E_\gamma\ge E_0$ on photon energy is performed. For values of 
$E_0$ below about 1.9\,GeV, the rate can be calculated without reference to 
shape functions using a multi-scale operator product expansion (MSOPE). The 
transition from the shape-function region to the MSOPE region is studied 
analytically. The resulting prediction for the $B\to X_s\gamma$ branching 
ratio depends on three large scales: $m_b$, $\sqrt{m_b\Delta}$, and 
$\Delta=m_b-2E_0$. Logarithms associated with these scales are resummed 
at next-to-next-to-leading logarithmic order. While power corrections in
$\Lambda_{\rm QCD}/\Delta$ turn out to be small, the sensitivity to the scale 
$\Delta\approx 1.1$\,GeV (for $E_0\approx 1.8$\,GeV) introduces significant 
perturbative uncertainties, which so far have been ignored. The new 
theoretical prediction for the $B\to X_s\gamma$ branching ratio with 
$E_\gamma\ge 1.8$\,GeV is $\mbox{Br}(B\to X_s\gamma)%
=(3.38_{\,-0.42\,-0.30}^{\,+0.31\,+0.32})\times 10^{-4}$, where the first 
error is an estimate of perturbative uncertainties and the second one reflects 
uncertainties in input parameters. With this cut $(89_{\,-7}^{\,+6}\,\pm 1)\%$ 
of all events are contained. The implications of larger theory uncertainties 
for New Physics searches are briefly explored with the example of the type-II 
two-Higgs-doublet model, for which the lower bound on the charged-Higgs mass 
is reduced compared with previous estimates to approximately 200\,GeV at 95\% 
confidence level.
\end{abstract}
\vfil

\end{titlepage}

\section{Introduction}

The inclusive, weak radiative decay $B\to X_s\gamma$ is the prototype of all 
flavor-changing neutral current processes. In the Standard Model, this process 
is mediated by loop diagrams containing $W$ bosons and top (or lighter) 
quarks. In extensions of the Standard Model, other heavy particles propagating 
in loops can give sizable contributions, which in many cases can compete with 
those of the Standard Model. As a result, measurements of the $B\to X_s\gamma$ 
rate and CP asymmetry provide sensitive probes for New Physics at the TeV 
scale. In many cases, the fact that these measurements agree with Standard 
Model predictions imposes non-trivial constraints on the allowed parameter 
space.

Given the prominent role of $B\to X_s\gamma$ decay in searching for physics 
beyond the Standard Model, it is of great importance to have a precise 
prediction for its inclusive rate and CP asymmetry in the Standard Model. This 
has been achieved thanks to the combined effort of many theorists over a 
period of several years \cite{Hurth:2003vb}. The total inclusive rate is known 
at next-to-leading order in renormalization-group (RG) improved perturbation 
theory with a theoretical precision of about 10\%. Currently, a major effort 
is underway to improve this accuracy by calculating the dominant parts of the 
next-to-next-to-leading corrections \cite{Bieri:2003ue,Misiak:2004ew}.

While the total inclusive $B\to X_s\gamma$ decay rate can be calculated using 
a conventional operator-product expansion (OPE) based on an expansion in 
logarithms and inverse powers of the $b$-quark mass \cite{Falk:1993dh}, the 
situation is more complicated when a cut on the photon energy is applied. In 
practice, experiments can only measure the high-energy part of the photon 
spectrum, $E_\gamma\ge E_0$, where typically $E_0=2$\,GeV (measured in the 
$B$-meson rest frame) or slightly lower \cite{Chen:2001fj,Koppenburg:2004fz}. 
Even if such a cut was not required for experimental reasons, it would be 
needed to reduce the photon background from $B\to X_s\psi^{(\prime)}$ decays 
followed by a radiative decay of the $\psi^{(\prime)}$ \cite{Gambino:2001ew}.
With $E_\gamma$ restricted to be close to the kinematic endpoint at $M_B/2$ 
(neglecting the kaon mass), the hadronic final state $X_s$ is constrained to 
have large energy $E_X\sim M_B$ but only moderate invariant mass 
$M_X\sim(M_B\Lambda_{\rm QCD})^{1/2}$. In this kinematic region, important 
hadronic effects need to be taken into account. An infinite set of 
leading-twist terms in the OPE need to be resummed into a non-perturbative 
shape function, which describes the momentum distribution of the $b$-quark 
inside the $B$ meson \cite{Neubert:1993ch,Bigi:1993ex,Mannel:1994pm}. In 
addition, Sudakov double logarithms arise near the endpoint of the photon 
spectrum, which need to be resummed to all orders in perturbation theory 
\cite{Falk:1993vb,Korchemsky:1994jb,Akhoury:1995fp}. While these issues are 
now well understood theoretically \cite{Bosch:2004th,Bauer:2003pi}, the 
presence of the shape function leads to an unavoidable element of hadronic 
uncertainty and modeling, which is undesirable when the goal is to probe for 
physics beyond the Standard Model.

Conventional wisdom says that, while shape-function effects are important near 
the endpoint of the photon spectrum, these effects can be ignored as soon as 
the cutoff $E_0$ is lowered below about 1.9\,GeV. This assumption is based on 
phenomenological studies of shape-function effects using various model 
functions, which have the unrealistic feature that the distribution 
function vanishes exponentially for large light-cone momenta
\cite{Dikeman:1995ad,Kagan:1998ym}. In other words, it has been implicitly 
assumed that there is an instantaneous transition from the ``shape-function 
region'' of large non-perturbative corrections to the ``OPE region'', in which 
hadronic corrections to the rate are suppressed by at least two powers of 
$\Lambda_{\rm QCD}/m_b$. As a result, the preferred strategy has 
been to encourage experimenters to lower the photon-energy cut to a value 
$E_0\le 1.9$\,GeV, and then to employ the conventional OPE for the calculation 
of the rate, ignoring shape-function effects.

In this paper we show that this strategy is based on a misconception. Our work 
is motivated by two considerations. First, is has recently been shown that the 
asymptotic behavior of $B$-meson distribution functions such as the shape 
function is not exponential, but rather governed by radiative tails exhibiting 
a slow, power-like fall-off \cite{Bosch:2004th,Lange:2003ff}. One should 
therefore not exclude the possibility of a significant radiation tail in the 
case of the $B\to X_s\gamma$ photon spectrum, meaning that more events than 
predicted by existing models could be located at low photon energy. Fits to 
experimental data in the low-energy part of the spectrum, which are based on 
such models, should thus be taken with caution. Secondly, it has been our 
desire for a long time to find an analytic way to study the transition from 
the shape-function region to the OPE region. If it were true that 
shape-function effects become irrelevant once the cutoff $E_0$ is lowered 
below 1.9\,GeV, one should be able to see this analytically using some form of 
a short-distance expansion. We show that this expansion indeed exists, and 
that it involves three different short-distance scales. In addition to the 
hard scale $m_b$, an intermediate ``hard-collinear'' scale $\sqrt{m_b\Delta}$ 
corresponding to the typical invariant mass of the hadronic final state $X_s$, 
and a low scale $\Delta=m_b-2E_0$ related to the width of the energy window 
over which the measurement is performed, become of crucial importance. The 
physics associated with these scales can be disentangled using recent results 
on soft-collinear factorization theorems derived in the framework of effective 
field theory \cite{Bosch:2004th,Bauer:2003pi}. A systematic treatment consists 
of matching QCD onto soft-collinear effective theory (SCET) 
\cite{Bauer:2000yr} in a first step, in which hard quantum fluctuations are 
integrated out. In a second step, hard-collinear modes are integrated out by 
matching SCET onto heavy-quark effective theory (HQET) \cite{Neubert:1993mb}. 
Ultimately, the precision of the theoretical calculations is determined by the 
value of the lowest short-distance scale $\Delta$, which in practice is of 
order 1\,GeV or only slightly larger. The theoretical accuracy that can be 
reached is therefore not as good as in the case of a conventional heavy-quark 
expansion applied to the $B$ system, but more likely it is similar to (if not 
worse than) the accuracy reached in the description of the inclusive hadronic 
$\tau$ decay rate $R_\tau$ \cite{Braaten:1991qm}. However, while the ratio 
$R_\tau$ is known to order $\alpha_s^3$, the $B\to X_s\gamma$ branching ratio 
is currently only known through order $\alpha_s$.

While we are aware that this conclusion may come as a surprise to many 
practitioners in the field of flavor physics, we believe that it is an 
unavoidable consequence of the analysis presented in this paper. Not 
surprisingly, then, we find that the error estimates for the partial 
$B\to X_s\gamma$ branching ratio in the literature are too optimistic. Since 
there are unknown $\alpha_s^2(\Delta)$ corrections at the low scale 
$\Delta\sim 1$\,GeV, we estimate the present perturbative uncertainty in the 
$B\to X_s\gamma$ branching ratio with $E_0$ in the range between 1.6 and 
1.8\,GeV to be of order 10\%. In addition, there are uncertainties due to 
other sources, such as the $b$- and $c$-quark masses. The combined theoretical 
uncertainty is of order 15\%. While this is a rather pessimistic conclusion, 
we stress that the uncertainty is limited by unknown, higher-order 
perturbative terms, not by non-perturbative effects, which we find to be under 
good control. (This is similar to the case of $R_\tau$.) Therefore, there is 
room for a reduction of the error by means of well-controlled perturbative 
calculations.

In Section~\ref{sec:fact}, we discuss the QCD factorization formula for the
partial $B\to X_s\gamma$ decay rate with a cut $E_\gamma\ge E_0$ on photon 
energy, valid at leading power in the heavy-quark expansion. Contributions 
associated with the hard, hard-collinear, and soft scales are factorized into 
a hard function $H_\gamma$, a jet function $J$, and a shape function $\hat S$. 
Single and double (Sudakov) logarithms are systematically resummed to all 
orders in perturbation theory. The RG evolution of the shape function is 
studied in Section~\ref{sec:SFRG}, where we present the exact solution to its 
evolution equation in momentum space. Our main results are derived in 
Section~\ref{sec:newOPE}, where we show how the convolution integral over the 
shape function in the factorization formula can be calculated using a local 
OPE, provided that the scale $\Delta=m_b-2E_0$ is numerically large compared 
with $\Lambda_{\rm QCD}$. Section~\ref{sec:scheme} discusses how to eliminate 
the HQET parameters $m_b$ and $\lambda_1$ defined in the pole scheme, which 
enter the theoretical expressions, in terms of physical parameters defined in 
the so-called ``shape-function scheme'' \cite{Bosch:2004th}. The calculation 
of the decay rate is completed in Section~\ref{sec:power}, where we add  
contributions that are power-suppressed in the heavy-quark expansion. For 
these small corrections, the scale separation we achieve is only approximate 
and misses some yet unknown terms of order $\alpha_s^2\ln^2(\Delta/m_b)$. In 
Section~\ref{sec:ratios}, we show that by considering ratios of decay rates 
one may separate the short-distance physics contained in the hard function 
$H_\gamma$ from the physics associated with the intermediate and low scales. 
For instance, at leading power in $\Delta/m_b$ the ratio of the 
$B\to X_s\gamma$ branching ratio in a New Physics model relative to that in 
the Standard Model can be calculated without any sensitivity to scales less 
than the hard scale $\mu_h\sim m_b$, and the same is true for the direct CP 
asymmetry. In other cases, at leading power some ratios are insensitive to 
the Wilson coefficients in the effective weak Hamiltonian and thus to New 
Physics. Examples are the average photon energy $\langle E_\gamma\rangle$, and 
the ratio of the $B\to X_s\gamma$ decay rate with $E_\gamma\ge E_0$ normalized 
to the total rate. The latter ratio is particularly interesting, since it can 
be used to make contact between a simple, fully inclusive rate calculation and 
our more sophisticated analysis of multi-scale effects. Our numerical results 
are presented in Section~\ref{sec:numerics}, followed by a summary and 
conclusions.

\section{QCD factorization theorem}
\label{sec:fact}

Recent results on the factorization of hard, hard-collinear, and soft 
contributions to inclusive $B$-meson decay distributions 
\cite{Bosch:2004th,Bauer:2003pi} allow us to obtain a QCD factorization 
formula for the integrated $B\to X_s\gamma$ decay rate with a cut 
$E_\gamma\ge E_0$ on photon energy. In the region of large $E_0$, the leading 
contribution to the rate can be written in the form
\begin{eqnarray}\label{ff}
   \Gamma_{\bar B\to X_s\gamma}(E_0)
   &=& \frac{G_F^2\alpha}{32\pi^4}\,|V_{tb} V_{ts}^*|^2\,
    (1+\varepsilon_{\rm np})\,
    \overline{m}_b^2(\mu_h)\,|H_\gamma(\mu_h)|^2\,U_1(\mu_h,\mu_i) \\
   &&\hspace{-1.9cm}\times \int_0^{\Delta_E}\!dP_+\,(M_B-P_+)^3
    \int_0^{P_+}\!d\hat\omega\,m_b\,J\big(m_b(P_+-\hat\omega),\mu_i\big)\,
    \hat S(\hat\omega,\mu_i) + \mbox{power corrections,} \nonumber
\end{eqnarray}
where $\Delta_E=M_B-2E_0$ is twice the width of the window in photon energy 
over which the measurement of the decay rate is performed. In the prefactor, 
$\alpha$ is the fine-structure constant normalized at $q^2=0$ 
\cite{Czarnecki:1998tn}. The variable $P_+=E_X-|\vec{P}_X|$ is the ``plus 
component'' of the 4-momentum of the hadronic final state $X_s$, which is 
related to the photon energy by $P_+=M_B-2E_\gamma$. The factor $(M_B-P_+)^3$ 
under the integral thus equals $8E_\gamma^3$, where two powers of $E_\gamma$ 
come from the squared matrix element of the effective weak Hamiltonian, and 
one factor comes from phase space. The hadronic invariant mass of the final 
state is $M_X^2=M_B P_+$. The endpoint region of the photon spectrum is 
defined by the requirement that $P_+\le\Delta_E\ll M_B$, in which case $P^\mu$ 
is called a hard-collinear momentum \cite{Bosch:2003fc}. Power corrections to 
the expression above will be analyzed later; however, the leading 
non-perturbative corrections to the total decay rate have already been 
factored out in (\ref{ff}) and included in the parameter \cite{Falk:1993dh} 
(see Table~\ref{tab:inputs} in Section~\ref{sec:numerics} for a list of input 
parameters)
\begin{equation}\label{epsnp}
   \varepsilon_{\rm np} = \frac{\lambda_1-9\lambda_2}{2m_b^2}
   = - (3.1\pm 0.5)\% \,.
\end{equation}

The factorization formula (\ref{ff}) was first presented in 
\cite{Korchemsky:1994jb}. What is new is that we now have a systematic 
effective field-theory technology to compute the functions $H_\gamma$ and $J$ 
order by order in perturbation theory, and to control their scale dependence 
in momentum space (not moment space). Also, it is in principle possible to 
include power-suppressed terms in the heavy-quark expansion. In the 
factorization formula, $\mu_h\sim m_b$ is a hard scale, while 
$\mu_i\sim\sqrt{m_b\Lambda_{\rm QCD}}$ is an intermediate hard-collinear scale 
of order the invariant mass of the hadronic final state. The precise values of 
these matching scales are irrelevant, since the rate is formally independent 
of $\mu_h$ and $\mu_i$. The hard corrections captured by the function 
$H_\gamma(\mu_h)$ result from the matching of the effective weak Hamiltonian 
of the Standard Model (or any of its extensions) onto a leading-order current 
operator of SCET. It is defined by the relation
\begin{equation}\label{match}
   {\cal H}_{\rm eff}^{b\to s\gamma}
   \to \frac{G_F}{\sqrt2}\,V_{tb} V_{ts}^*\,\frac{e}{2\pi^2}\,E_\gamma\,
   \overline{m}_b(\mu_h)\,H_\gamma(\mu_h)\,\epsilon_\mu^*(q)\,
   \left[ \bar\xi\,W_{hc}\,\gamma_\perp^\mu (1-\gamma_5)\,h_v \right]\!(\mu_h)
   + \dots \,,
\end{equation}
where $\epsilon(q)$ is the transverse photon polarization vector, and the dots 
represent power-sup\-pressed contributions from higher-dimensional SCET 
operators. The result is proportional to the photon energy, 
$E_\gamma=v\cdot q$, defined in the $B$-meson rest frame. (Here $v$ is the 
4-velocity of the $B$ meson.) The fields $h_v$ and $\xi$ represent the soft 
heavy quark and the hard-collinear strange quark, respectively, and $W_{hc}$ 
is a Wilson line. At tree level, only the dipole operator $Q_{7\gamma}$ and 
the four-quark penguin operators $Q_5$ and $Q_6$ in the effective weak 
Hamiltonian give a non-zero contribution to $H_\gamma$, which is equal to the 
``effective'' coefficient 
$C_{7\gamma}^{\rm eff}=C_{7\gamma}-\frac13\,C_5-C_6$. (We use the conventions 
of \cite{Beneke:2001ev} for the operators and Wilson coefficients in the 
effective weak Hamiltonian.) At next-to-leading order, the result reads (with 
$C_F=4/3$)
\begin{eqnarray}\label{Hgamma}
   H_\gamma(\mu_h)
   &=& C_{7\gamma}^{\rm eff}(\mu_h) \left[ 1
    + \frac{C_F\alpha_s(\mu_h)}{4\pi} \left( -2\ln^2\frac{m_b}{\mu_h}
    + 7\ln\frac{m_b}{\mu_h} - 6 - \frac{\pi^2}{12} \right) 
    + \varepsilon_{\rm ew} \right] \nonumber\\
   &&\mbox{}+ C_{8g}^{\rm eff}(\mu_h)\,\frac{C_F\alpha_s(\mu_h)}{4\pi}
    \left( - \frac83 \ln\frac{m_b}{\mu_h} + \frac{11}{3} - \frac{2\pi^2}{9}
    + \frac{2\pi i}{3} \right) \nonumber\\
   &&\mbox{}+ C_1(\mu_h)\,\frac{C_F\alpha_s(\mu_h)}{4\pi}
    \left( \frac{104}{27} \ln\frac{m_b}{\mu_h} + g(z)
    + \varepsilon_{\rm CKM}\,\big[ g(0) - g(z) \big] \right)
    + \varepsilon_{\rm peng} \,.
\end{eqnarray}
The coefficient $C_{7\gamma}^{\rm eff}(\mu_h)$ of the electromagnetic dipole 
operator is required with next-to-leading order accuracy 
\cite{Chetyrkin:1996vx}, while the remaining coefficients can be calculated at 
leading logarithmic order. Explicit expressions for these coefficients can be
found, e.g., in \cite{Kagan:1998ym,Chetyrkin:1996vx}. The terms in the third 
row arise from charm-quark and up-quark penguin contractions of the 
current-current operators $Q_1^{c,u}$. These contributions depend on the small 
ratio 
\begin{equation}\label{epsCKM}
   \varepsilon_{\rm CKM}
   = - \frac{V_{ub} V_{us}^*}{V_{tb} V_{ts}^*}
   = \lambda^2(\bar\rho-i\bar\eta) \left[ 1 + \lambda^2(1-\bar\rho-i\bar\eta)
   + {\cal O}(\lambda^4) \right] .
\end{equation}
The variable $z=(m_c/m_b)^2$ denotes the ratio of quark masses relevant to the 
charm loop, and
\begin{eqnarray}\label{gz}
   g(z) &=& - \frac{833}{162} - \frac{20\pi i}{27} + \frac{8\pi^2}{9}\,z^{3/2}
    \nonumber\\
   &&\mbox{}+ \frac{2z}{9} \left[ 48 - 5\pi^2\! - 36\zeta_3
    + (30\pi - 2\pi^3) i + \!(36 - 9\pi^2 + 6\pi i) \ln z
    + \!(3 + 6\pi i) \ln^2 z + \ln^3 z \right] \nonumber\\
   &&\mbox{}+ \frac{2z^2}{9} \left[ 18 + 2\pi^2 - 2\pi^3 i
    + (12 - 6\pi^2) \ln z + 6\pi i\ln^2 z + \ln^3 z \right] \nonumber\\
   &&\mbox{}+ \frac{z^3}{27} \left[ -9 - 14\pi^2 + 112\pi i
    + (182 - 48\pi i) \ln z - 126\ln^2 z \right] + \dots
\end{eqnarray}
are the first few terms in the expansion of the penguin function 
\cite{Greub:1996tg}, whose exact expression (in the form of parameter 
integrals) can be found in \cite{Buras:2002tp}. The imaginary parts in 
(\ref{Hgamma}) and (\ref{gz}) are strong-interaction phases, which in 
conjunction with CP-violating weak phases contained in the parameter 
$\varepsilon_{\rm CKM}$ or in potential New Physics contributions to the 
Wilson coefficients can induce a non-zero CP asymmetry in $B\to X_s\gamma$ 
decays \cite{Soares:1991te,Kagan:1998bh}. The term
\begin{equation}
   \varepsilon_{\rm ew} = \delta_{\rm ew}
   + \frac{\alpha(\mu_h)}{\alpha_s(\mu_h)}\,
   \frac{C_{7\gamma}^{\rm (em)}(\mu_h)}{C_{7\gamma}^{\rm eff}(\mu_h)} 
   \approx - 1.5\% 
\end{equation}
accounts for electroweak matching corrections at the weak scale 
\cite{Gambino:2000fz} and logarithmically enhanced electromagnetic effects 
affecting the evolution of the Wilson coefficients 
\cite{Kagan:1998ym,Baranowski:1999tq}. Finally, the term 
$\varepsilon_{\rm peng}\approx 0.2\%$ includes the effects of penguin 
contractions of operators other than $Q_1^{c,u}$ \cite{Buras:2002tp}, which 
are numerically negligible but are included here for completeness. In the 
factorization formula (\ref{ff}), the hard function is multiplied by the 
running $b$-quark mass 
\begin{equation}
   \overline{m}_b(\mu_h) = \overline{m}_b(\overline{m}_b) \left[
   1 + \frac{3C_F\alpha_s(\mu_h)}{2\pi}\,
   \ln\frac{\overline{m}_b}{\mu_h} + \dots \right]
\end{equation}
defined in the $\overline{\rm MS}$ scheme, which is part of the 
electromagnetic dipole operator $Q_{7\gamma}$. On the other hand, the scheme 
to be used for the quark masses entering the ratio $z$ in the penguin function 
$g(z)$ is not specified at next-to-leading order \cite{Gambino:2001ew}. Since 
the matching is performed at a hard scale $\mu_h$, the charm-quark mass should 
be a running mass $\overline{m}_c(\mu_h)$, while $m_b$ enters either as the 
mass in the $b$-quark propagator or via the values of external momenta. 
For simplicity, we take $z=[\overline{m}_c(\mu_h))/\overline{m}_b(\mu_h)]^2$ 
as a ratio of running quark masses evaluated at the same scale, which has the 
advantage that this quantity is RG invariant.

\newpage 
The jet function $J\big(m_b(P_+-\hat\omega),\mu_i\big)$ in (\ref{ff}) 
describes the physics of the final-state hadronic jet. At next-to-leading 
order in perturbation theory, it is given by the expression 
\cite{Bosch:2004th,Bauer:2003pi}
\begin{eqnarray}\label{Jres}
   m_b\,J\big(m_b(P_+-\hat\omega),\mu_i\big)
   &=& \delta(P_+ - \hat\omega) \left[ 1 + \frac{C_F\alpha_s(\mu_i)}{4\pi}
    \left( 7 - \pi^2 \right) \right] \nonumber\\
   &&\mbox{}+ \frac{C_F\alpha_s(\mu_i)}{4\pi} \left[
   \frac{1}{P_+ - \hat\omega} \left( 4\ln\frac{m_b(P_+ - \hat\omega)}{\mu_i^2}
   - 3 \right) \right]_*^{[\mu_i^2/m_b]}\! . ~
\end{eqnarray}
The star distributions are generalized plus distributions defined as 
\cite{DeFazio:1999sv}
\begin{eqnarray}
   \int_{\le 0}^z\!dx\,F(x) 
   \left[ \frac{1}{x} \right]_*^{[u]}
   &=& \int_0^z\!dx\,\frac{F(x)-F(0)}{x} + F(0)\,\ln\frac{z}{u} \,, \nonumber\\
   \int_{\le 0}^z\!dx\,F(x)
    \left[ \frac{\ln(x/u)}{x} \right]_*^{[u]}
   &=& \int_0^z\!dx\,\frac{F(x)-F(0)}{x}\,\ln\frac{x}{u} 
    + \frac{F(0)}{2}\,\ln^2\frac{z}{u} \,,
\end{eqnarray}
where $F(x)$ is a smooth test function. The perturbative expansion of the jet 
function can be trusted as long as $\mu_i^2\sim m_b\Delta$ with 
$\Delta\sim P_+^{\rm max}-\langle\hat\omega\rangle\simeq m_b-2E_0\ll M_B$. By 
quark-hadron duality, only the maximum values of kinematic variables such as 
$P_+$, which are integrated over phase space, matter for the calculation of 
inclusive decay rates \cite{Falk:1993dh}. Note that the ``natural'' choices 
$\mu_h\propto m_b$ and $\mu_i^2\equiv m_b\,\tilde\mu_i$ with $\tilde\mu_i$ 
independent of $m_b$ remove all reference to the $b$-quark mass (other than in 
the arguments of running coupling constants) from the factorization formula 
(\ref{ff}).

The shape function $\hat S(\hat\omega,\mu_i)$ parameterizes our ignorance 
about the soft physics associated with bound-state effects inside the $B$ 
meson \cite{Neubert:1993ch,Bigi:1993ex}. Its naive interpretation is that of a 
parton distribution function, governing the distribution of the light-cone 
component $k_+$ of the residual momentum $k=p_b-m_b v$ of the $b$ quark inside 
the heavy meson. Once radiative corrections are included, however, a 
probabilistic interpretation of the shape function breaks down 
\cite{Bosch:2004th}. For convenience, the shape function is renormalized in 
(\ref{ff}) at the intermediate hard-collinear scale $\mu_i$ rather than at a 
hadronic scale $\mu_{\rm had}$. This removes any uncertainties related to the 
evolution from $\mu_i$ to $\mu_{\rm had}$. Since the shape function is 
universal, all that matters is that it is renormalized at the same scale when 
comparing different processes.

The last ingredient in the factorization formula (\ref{ff}) is the RG 
evolution function $U_1(\mu_h,\mu_i)$, which describes the evolution of the 
hard function $|H_\gamma|^2$ from the high matching scale $\mu_h$ down to the 
intermediate scale $\mu_i$, at which the jet and shape functions are 
renormalized. The exact expression for this quantity follows from
\begin{equation}\label{U1}
   \ln U_1(\mu_h,\mu_i)
   = 2S(\mu_h,\mu_i) - 2a_\Gamma(\mu_h,\mu_i) \ln\frac{m_b}{\mu_h}
   - 2a_{\gamma'}(\mu_h,\mu_i) \,,
\end{equation}
where the various functions on the right-hand side are the solutions to the 
partial differential equations (the Sudakov exponent $S$ should not be 
confused with the shape function $\hat S$)
\begin{eqnarray}\label{dgl}
   \frac{d}{d\ln\mu}\,S(\nu,\mu)
   &=& - \Gamma_{\rm cusp}\big(\alpha_s(\mu)\big)\,\ln\frac{\mu}{\nu} \,,
    \nonumber\\
   \frac{d}{d\ln\mu}\,a_\Gamma(\nu,\mu)
   &=& - \Gamma_{\rm cusp}\big(\alpha_s(\mu)\big) \,, \qquad
   \frac{d}{d\ln\mu}\,a_{\gamma'}(\nu,\mu)
    = - \gamma'\big(\alpha_s(\mu)\big) \,,
\end{eqnarray}
with initial conditions $S(\nu,\nu)=a_\Gamma(\nu,\nu)=a_{\gamma'}(\nu,\nu)=0$ 
at $\mu=\nu$. Here $\Gamma_{\rm cusp}$ is the universal cusp anomalous 
dimension for Wilson loops with light-like segments \cite{Korchemsky:wg}, 
which has recently been calculated to three-loop order \cite{Moch:2004pa}, and 
$\gamma'$ enters the anomalous dimension of the leading-order SCET current 
operators containing a heavy quark and a hard-collinear quark with large 
energy $E$, which takes the form \cite{Bauer:2000yr,Bosch:2003fc}
\begin{equation}\label{gamJ}
   \gamma_J(E,\mu) = - \Gamma_{\rm cusp}\big(\alpha_s(\mu)\big) 
   \ln\frac{\mu}{2E} + \gamma'\big(\alpha_s(\mu)\big) \,.
\end{equation}
As explained in Appendix~A.1, a conjecture for the two-loop expression for 
$\gamma'$ can be deduced using results from the literature on deep-inelastic 
scattering \cite{Vogt:2000ci,Gardi:2004ia,inprep}. The 
evolution equations (\ref{dgl}) are solved in the standard way by writing 
$d/d\ln\mu=\beta(\alpha_s)\,d/d\alpha_s$, where 
$\beta(\alpha_s)=d\alpha_s/d\ln\mu$ is the QCD $\beta$ function. This yields 
the exact solutions
\begin{equation}\label{RGEsols}
   S(\nu,\mu) = - \int\limits_{\alpha_s(\nu)}^{\alpha_s(\mu)}\!
    d\alpha\,\frac{\Gamma_{\rm cusp}(\alpha)}{\beta(\alpha)}
    \int\limits_{\alpha_s(\nu)}^\alpha
    \frac{d\alpha'}{\beta(\alpha')} \,, \qquad
   a_\Gamma(\nu,\mu) = - \int\limits_{\alpha_s(\nu)}^{\alpha_s(\mu)}\!
    d\alpha\,\frac{\Gamma_{\rm cusp}(\alpha)}{\beta(\alpha)} \,, 
\end{equation}
and similarly for the function $a_{\gamma'}$. The perturbative expansions of 
the anomalous dimensions and the resulting expressions for the evolution 
functions valid through order $\alpha_s$ are collected in Appendix~A.1. 

\begin{figure}
\begin{center}
\epsfig{file=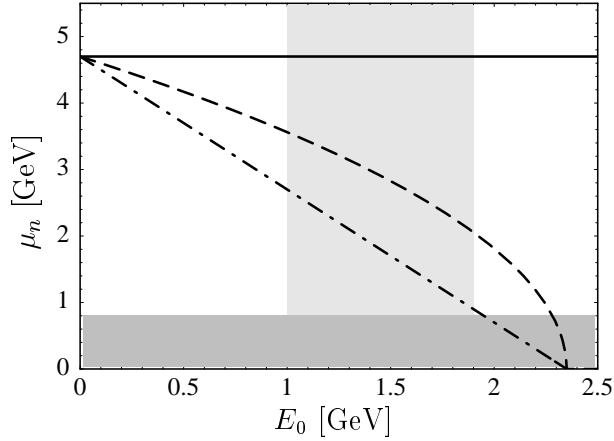,width=8cm}
\end{center}
\vspace{-0.4cm}
\centerline{\parbox{14cm}{\caption{\label{fig:scales}
Dependence of the three scales $\mu_h=m_b$ (solid), $\mu_i=\sqrt{m_b\Delta}$ 
(dashed), and $\mu_0=\Delta$ (dash-dotted) on the cutoff $E_0$, assuming 
$m_b=4.7$\,GeV. The gray area at the bottom shows the domain of 
non-perturbative physics. The light gray band in the center indicates the 
region where the MSOPE should be applied.}}}
\end{figure}

As written in (\ref{ff}), the decay rate is sensitive to non-perturbative
hadronic physics via its dependence on the shape function. This sensitivity is 
unavoidable as long as the scale $\Delta=m_b-2E_0$ is a hadronic scale,
corresponding to the endpoint region of the photon spectrum above, say, 
2\,GeV. The properties of the $B\to X_s\gamma$ decay rate and photon spectrum 
in this region will be discussed in detail elsewhere. Here we are 
interested in a situation where $E_0$ is lowered out of the shape-function 
region, such that $\Delta$ can be considered large compared with 
$\Lambda_{\rm QCD}$. For orientation, we note that with $m_b=4.7$\,GeV and the 
cutoff $E_0=1.8$\,GeV employed in a recent analysis by the Belle Collaboration 
\cite{Koppenburg:2004fz} one gets $\Delta=1.1$\,GeV. For $E_0=1.6$\,GeV (a 
reference value adopted in \cite{Gambino:2001ew,Buras:2002tp}, which at 
present is below what can be achieved experimentally) one would obtain 
$\Delta=1.5$\,GeV. As mentioned in the Introduction, in all previous analyses 
of the $B\to X_s\gamma$ decay rate it was assumed that, once $E_0$ is taken 
below about 1.9\,GeV, the sensitivity to hadronic physics essentially 
disappears, 
and the rate can be computed using a conventional OPE at the scale $m_b$. The 
main point of the present work is to show that this assumption cannot be 
justified, and that estimating theoretical uncertainties under the hypothesis 
that the expansion is in powers of $\alpha_s(m_b)$ and $\Lambda_{\rm QCD}/m_b$ 
underestimates the magnitude of the true theoretical errors. As we will show, 
for values of $E_0$ outside the shape-function region there are three relevant 
mass scales in the problem besides $\Lambda_{\rm QCD}$. They are the hard 
scale $m_b$, the hard-collinear scale $\sqrt{m_b\Delta}$, and the low scale 
$\Delta$ itself. The values of these scales as a function of the photon-energy 
cutoff $E_0$ are shown in Figure~\ref{fig:scales}. The transition from the 
shape-function region to the region where a conventional OPE can be applied is 
not abrupt but proceeds  via an intermediate region, in which a short-distance 
analysis based on a multi-stage OPE (MSOPE) can be performed. The transition 
from the shape-function region into the MSOPE region occurs when the scale 
$\Delta$ becomes numerically (but not parametrically) large compared with 
$\Lambda_{\rm QCD}$. Then terms of order $\alpha_s^n(\Delta)$ and 
$(\Lambda_{\rm QCD}/\Delta)^n$, which are non-perturbative in the 
shape-function region, gradually become decent expansion parameters. Only for 
very low values of the cutoff ($E_0<1$\,GeV or so) it is justified to treat 
$\Delta$ and $\sqrt{m_b\Delta}$ as scales of order $m_b$. 

Separating the contributions associated with these scales requires a 
multi-step procedure, which we develop in the present work. The
first step, the separation of the hard scale from the intermediate scale, has
already been achieved in (\ref{ff}). To proceed further we use two crucial 
recent developments. First, integrals of smooth weight functions 
$F(\hat\omega)$ with the shape function $\hat S(\hat\omega,\mu)$ can be 
expanded in a series of forward $B$-meson matrix elements of local HQET 
operators, provided that the integration domain is large compared with 
$\Lambda_{\rm QCD}$ \cite{Bosch:2004th,Bauer:2003pi}. The reason is that the 
shape function can be written as the discontinuity of a two-point correlator 
in momentum space, and thus weighted integrals over $\hat S$ can be turned 
into contour integrals in the complex plane along a circle with radius set by 
the upper integration limit on $\hat\omega$ (more precisely, 
$\hat\omega-\bar\Lambda$). Specifically, the expansion takes the form
\cite{Bosch:2004th}
\begin{equation}\label{OPE}
   \int_0^{\Delta+\bar\Lambda}\!d\hat\omega\,\hat S(\hat\omega,\mu)\,
   F(\hat\omega) = K_0^{(F)}(\Delta,\mu)
   + K_2^{(F)}(\Delta,\mu)\,\frac{(-\lambda_1)}{3\Delta^2} + \dots \,, 
\end{equation}
where $K_n^{(F)}$ are calculable Wilson coefficient functions, 
$\bar\Lambda=m_B-m_b$ and $\lambda_1$ are HQET parameters (which for the time 
being are defined in the pole scheme) \cite{Neubert:1993mb}, and the dots 
represent terms of order $(\Lambda_{\rm QCD}/\Delta)^3$ or higher. Note that 
with $\Delta=m_b-2E_0$ as defined above we have 
$\Delta+\bar\Lambda=M_B-2E_0=\Delta_E$, which coincides with the upper limit 
for the integration over $\hat\omega$ in (\ref{ff}). The perturbative 
expansions of the coefficient functions $K_n^{(F)}$ can be trusted as long as 
$\mu\sim\Delta$. In order to complete the scale separation, it is therefore 
necessary to evolve the shape function in (\ref{ff}) from the intermediate 
scale $\mu_i\sim\sqrt{m_b\Delta}$ down to a scale $\mu_0\sim\Delta$. This can 
be achieved using the analytic solution to the integro-differential RG 
evolution equation for the shape function in momentum space obtained in 
\cite{Bosch:2004th,Lange:2003ff}. These manipulations will be discussed in 
detail in the following two sections.

As a final comment, we stress that the main purpose of performing the scale
separation using the MSOPE is not that this allows us to resum Sudakov 
logarithms by solving RG equations. Indeed, the ``large logarithm''
$\ln(m_b/\Delta)\approx 1.5$ is only parametrically large, but not 
numerically. What is really important is to disentangle the physics at the low 
scale $\mu_0\sim\Delta$, which is ``barely perturbative'', from the physics 
associated with higher scales, where a short-distance treatment is on much 
safer grounds. It would be wrong to pretend that all perturbative effects in 
$B\to X_s\gamma$ decays are associated with the short-distance scale 
$m_b\gg\Lambda_{\rm QCD}$. The MSOPE allows us to distinguish between the 
three coupling constants $\alpha_s(m_b)\approx 0.22$, 
$\alpha_s(\sqrt{m_b\Delta})\approx 0.29$, and $\alpha_s(\Delta)\approx 0.44$
(for $\Delta=1.1$\,GeV), which are rather different despite the fact that 
there are no numerically large logarithms in the problem. Given the values of 
these couplings, we expect that scale separation between $\Delta$ and $m_b$ is 
as important as that between $m_b$ and the weak scale $M_W$.

\section{Evolution of the shape function}
\label{sec:SFRG}

The renormalized shape function obeys the integro-differential RG evolution 
equation
\begin{equation}\label{Sevol}
   \frac{d}{d\ln\mu}\,\hat S(\hat\omega,\mu)
   = - \int d\hat\omega'\,\gamma_S(\hat\omega,\hat\omega',\mu)\,
   \hat S(\hat \omega',\mu) \,,
\end{equation}
where the anomalous dimension can be written in the form
\begin{equation}\label{gammaS}
   \gamma_S(\hat\omega,\hat\omega',\mu)
   = - 2\Gamma_{\rm cusp}\big(\alpha_s(\mu)\big)
   \left[ \frac{1}{\hat\omega-\hat\omega'} \right]_*^{[\mu]}
   + 2\gamma\big(\alpha_s(\mu)\big)\,\delta(\hat\omega-\hat\omega') \,.
\end{equation}
This form was found in two recent one-loop calculations of the ultra-violet 
poles of non-local HQET operators \cite{Bosch:2004th,Bauer:2003pi}. A brief 
history of previous investigations of the anomalous-dimension kernel 
can be found in the first reference. The structure of (\ref{gammaS}) 
was derived first by Grozin and Korchemsky \cite{Grozin:1994ni}, who 
also computed the anomalous dimension and argued that the functional form of 
$\gamma_S(\hat\omega,\hat\omega',\mu)$ shown above holds to all orders in 
perturbation theory. A conjecture for the two-loop expression of the anomalous 
dimension $\gamma$ is presented in Appendix~A.1.

The exact solution to (\ref{Sevol}) can be found using a technique developed
in \cite{Lange:2003ff}. The equation is solved by the remarkably simple form
\begin{equation}\label{sonice}
   \hat S(\hat\omega,\mu_i) = U_2(\mu_i,\mu_0)\,
   \frac{e^{-\gamma_E\eta}}{\Gamma(\eta)} \int_0^{\hat\omega}\!d\hat\omega'\,
   \frac{\hat S(\hat\omega',\mu_0)}
        {\mu_0^\eta(\hat\omega-\hat\omega')^{1-\eta}} \,,
\end{equation}
where
\begin{equation}\label{U2}
   \ln U_2(\mu_i,\mu_0) = 2S(\mu_0,\mu_i) + 2a_\gamma(\mu_0,\mu_i) \,,
   \qquad
   \eta = -2a_\Gamma(\mu_0,\mu_i) \,.
\end{equation}
The functions $S$ and $a_\Gamma$ have been defined in (\ref{dgl}). Similarly, 
the function $a_\gamma$ is defined in complete analogy with $a_{\gamma'}$, but 
with $\gamma'$ replaced with the anomalous dimension $\gamma$ in 
(\ref{gammaS}). Explicit equations for these functions are given in 
Appendix~A.1. The next-to-leading logarithmic approximation to (\ref{sonice}) 
was first derived in \cite{Bosch:2004th}. We note that a similar (but not 
identical) result was found in \cite{{Balzereit:1998yf}} based on a one-loop 
calculation of the anomalous-dimension kernel.

Relation (\ref{sonice}) accomplishes the evolution of the shape function from 
the intermediate scale down to the low scale $\mu_0\sim\Delta$. When this 
result is inserted into the factorization formula (\ref{ff}), it is possible 
to perform the integrations over $P_+$ and $\hat\omega$ analytically, leaving 
the integration over $\hat\omega'$ until the end. Using the expression for the 
jet function in (\ref{Jres}), we find that the 
leading contribution to the decay rate is given by
\begin{equation}\label{rate2}
   \Gamma_{\bar B\to X_s\gamma}^{\rm leading}(E_0)
   = \frac{G_F^2\alpha}{32\pi^4}\,|V_{tb} V_{ts}^*|^2\,
   (1+\varepsilon_{\rm np})\,\overline{m}_b^2(\mu_h)\,|H_\gamma(\mu_h)|^2\,
   U_1(\mu_h,\mu_i)\,U_2(\mu_i,\mu_0)\,
   \frac{e^{-\gamma_E\eta}}{\Gamma(1+\eta)}\,I(E_0) \,,
\end{equation}
where
\begin{equation}\label{calIdef}
   I(E_0) = \int_0^{\Delta_E}\!d\hat\omega\, \hat S(\hat\omega,\mu_0)\,
   (M_B-\hat\omega)^3 \left( \frac{\Delta_E-\hat\omega}{\mu_0} \right)^\eta
   \left[ 1 + \frac{C_F\alpha_s(\mu_i)}{4\pi}\,{\cal J}(\Delta_E-\hat\omega)
   \right] p_3\Big(\eta,\frac{\Delta_E-\hat\omega}{M_B-\hat\omega}\Big) \,.
\end{equation}
The function $p_3(\eta,\delta)$ is a special case of the polynomial
\begin{equation}\label{fdef}
   p_n(\eta,\delta) = \sum_{k=0}^n
   \bigg( \begin{array}{c} n \\ k \end{array} \bigg)\,
   \frac{\eta\,(-\delta)^k}{k+\eta} \quad \Rightarrow \quad
   p_3(\eta,\delta) = 1 - \frac{3\eta\delta}{1+\eta}
   + \frac{3\eta\delta^2}{2+\eta} - \frac{\eta\delta^3}{3+\eta} \,.
\end{equation}
The next-to-leading order corrections from the jet function are encoded in the
operator
\begin{eqnarray}\label{calJ}
   {\cal J}(\Delta)
   &=& 2 \left( \ln\frac{m_b\Delta}{\mu_i^2} + \partial_\eta \right)^2
    - \big[ 4h(\eta) + 3 \big] \left( \ln\frac{m_b\Delta}{\mu_i^2}
    + \partial_\eta \right) \nonumber\\
   &&\mbox{}+ 2h^2(\eta) + 3h(\eta) - 2h'(\eta) + 7 - \frac{2\pi^2}{3} \,,
\end{eqnarray}
where
\begin{equation}
   h(\eta) = \psi(\eta) + \gamma_E + \frac{1}{\eta}
   = \psi(1+\eta) + \gamma_E
\end{equation}
is the harmonic function generalized to non-integer argument, and the 
derivatives $\partial_\eta=\partial/\partial\eta$ in (\ref{calJ}) act on the 
function $p_3(\eta,\delta)$ in (\ref{calIdef}). Note that this result has a 
smooth limit for $\mu_0\to\mu_i$, in which case $\eta\to 0$, 
$U_2(\mu_i,\mu_0)\to 1$, $h(\eta)\to 0$, $h'(\eta)\to\pi^2/6$, and we obtain 
an expression equivalent to the original result in (\ref{ff}).

\section{Short-distance expansion of the convolution integral}
\label{sec:newOPE}

The remaining task is to expand the integral over the shape function in 
(\ref{calIdef}) using an OPE, relating it to forward $B$-meson matrix elements
of local HQET operators, as indicated in (\ref{OPE}). As explained in 
\cite{Bosch:2004th}, this can be done whenever 
$\Delta=\Delta_E-\bar\Lambda=m_b-2E_0$ is large compared with 
$\Lambda_{\rm QCD}$. For a given weight function $F$, the matching 
coefficients $K_n^{(F)}$ are determined in the usual way by computing the 
integral in perturbation theory, expanding in powers of external momenta, and 
writing the answer as a linear combination of Wilson coefficients multiplying 
the matrix elements of local HQET operators. This matching calculation can be 
done using free partons in the external states and employing any infra-red 
regulator scheme that is convenient. We use on-shell external heavy-quark 
states with residual momentum $k$ chosen such that $v\cdot k=0$. In this case, 
the perturbative expression for the renormalized shape function at one-loop 
order is \cite{Bosch:2004th,Bauer:2003pi}
\begin{eqnarray}\label{Sparton}
   \hat S_{\rm parton}(\hat\omega,\mu_0)
   &=& \delta(\hat\omega-\bar\Lambda+n\cdot k)\,\left( 1
    - \frac{C_F\alpha_s(\mu_0)}{\pi}\,\frac{\pi^2}{24} \right) \nonumber\\
   &&\mbox{}- \frac{C_F\alpha_s(\mu_0)}{\pi}
    \left[ \frac{1}{\hat\omega-\bar\Lambda+n\cdot k}
    \left( 2\ln\frac{\hat\omega-\bar\Lambda+n\cdot k}{\mu_0} + 1 \right)
    \right]_*^{[\mu_0]} .
\end{eqnarray}
Using this result one can perform the integral over the shape function in 
(\ref{calIdef}) analytically. The answer is then Taylor-expanded in powers of
$n\cdot k$. The terms up to second order in this expansion are identified with 
the forward $B$-meson matrix elements of the operators $\bar h h$, 
$\bar h\,in\cdot D h$, and $\bar h\,(in\cdot D)^2 h$, respectively, where 
$n^\mu=(1,0,0,1)$ is a light-like vector. The values of these matrix elements 
are given by 1, 0, and $-\lambda_1/3$ \cite{Neubert:1993ch}. They do not 
receive radiative corrections in the regularization scheme adopted here. 
Operators of dimension six or higher would mix under renormalization. Also, in 
order to find their Wilson coefficients it would be necessary to perform 
matching calculations with external gluon states. However, it will be 
sufficient for all practical purposes to truncate the expansion after the 
second term, keeping only operators of dimension up to five. The result of 
this calculation is
\begin{eqnarray}\label{Sexpand}
   I(E_0) 
   &=& m_b^3 \left( \frac{\Delta}{\mu_0} \right)^\eta \left[
    1 + \frac{C_F\alpha_s(\mu_i)}{4\pi}\,{\cal J}(\Delta)
    + \frac{C_F\alpha_s(\mu_0)}{4\pi}\,{\cal S}(\Delta) \right]
    \nonumber\\
   &&\times \left[ p_3\Big(\eta,\frac{\Delta}{m_b}\Big)
    + \frac{\eta(\eta-1)}{2}\,\frac{(-\lambda_1)}{3\Delta^2} + \dots \right] , 
\end{eqnarray}
where ${\cal J}(\Delta)$ has been defined in (\ref{calJ}), and
\begin{equation}\label{calS}
   {\cal S}(\Delta)
   = - 4 \left( \ln\frac{\Delta}{\mu_0} + \partial_\eta \right)^2
   + 4 \left[ 2h(\eta) - 1 \right] \left( \ln\frac{\Delta}{\mu_0}
   + \partial_\eta \right) 
   - 4h^2(\eta) + 4h(\eta) + 4h'(\eta) - \frac{5\pi^2}{6} \,. \quad
\end{equation}
We have restricted ourselves to include only the leading power correction of 
order $\lambda_1/\Delta^2$, dropping terms that are suppressed by additional 
powers of $\Delta/m_b$. This is necessary for consistency, because there exist 
other, unknown $1/m_b$ and $1/m_b^2$ corrections from subleading shape 
functions, i.e., non-local HQET operators containing additional derivatives or 
insertions of soft gluon fields \cite{Bauer:2001mh}. The $\lambda_1/\Delta^2$ 
term is obtained by acting with $(-\lambda_1/6)\,\partial_\Delta^2$ on the 
leading-order term. According to (\ref{Sexpand}), its effect can be included 
by simply adding a power correction to the function $p_3(\eta,\delta)$.

The reader may ask why such an ``enhanced'' power correction was not found in 
previous analyses of the decay $B\to X_s\gamma$, or of the related 
semileptonic decay $B\to X_u\,l\,\nu$. Common lore is that non-perturbative 
corrections to inclusive decay rates scale like $(\Lambda_{\rm QCD}/m_b)^2$ 
and thus are very small. The reason is that so far power corrections in the 
OPE were computed at tree level only (an exception being 
\cite{Neubert:2001ib}). While the terms displayed above have a non-zero 
leading-order coefficient after RG resummation, they vanish at tree level if 
the result is expanded in fixed-order perturbation theory. Explicitly, we find 
to first order in $\alpha_s$
\begin{equation}
   \frac{I(E_0)}{m_b^3} 
   \ni \frac{(-\lambda_1)}{3\Delta^2}\,\frac{C_F\alpha_s}{4\pi}
   \left( -2\ln\frac{m_b}{\Delta} + \frac32 \right) .
\end{equation}
This effect would have shown up in the conventional heavy-quark expansion, if 
power corrections had been computed beyond the tree approximation.

\begin{figure}
\begin{center}
\epsfig{file=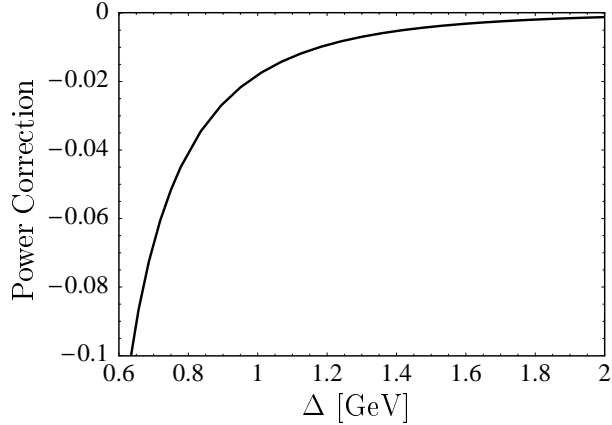,width=8cm}
\end{center}
\vspace{-0.4cm}
\centerline{\parbox{14cm}{\caption{\label{fig:powers}
Size of the enhanced power correction proportional to $\lambda_1/\Delta^2$ in
(\ref{Sexpand}) relative to the leading term, as a function of 
$\Delta=m_b-2E_0$.}}}
\end{figure}

Even though it is parametrically larger than the non-perturbative corrections 
from the conventional OPE in (\ref{epsnp}), the enhanced power correction in 
(\ref{Sexpand}) remains small for all relevant values of $\Delta$. This is 
illustrated in Figure~\ref{fig:powers}, which shows the size of the power 
correction proportional to $\lambda_1$ to the function $I(E_0)$ relative to 
the leading-order term. In the region of ``perturbative'' values of $\Delta$, 
where the MSOPE can be trusted, the effect amounts to a reduction of the decay 
rate by less than 5\%. It also follows that subleading power corrections of 
order $\lambda_1/(m_b\Delta)$ can safely be neglected.

Equation~(\ref{Sexpand}), combined with (\ref{rate2}), is our main result. Its 
numerical implications will be analyzed later, after including additional, 
small power-suppressed terms. A few comments are in order already at this 
point:
\begin{enumerate}
\item
The rate in (\ref{rate2}) is formally independent of the three matching 
scales, at which we switch from QCD to SCET ($\mu_h$), from SCET to HQET 
($\mu_i$), and finally at which the non-local HQET matrix element (the 
shape-function integral) is expanded in a series of local operators ($\mu_0$). 
The explicit perturbative expressions for the functions $H_\gamma(\mu_h)$ in
(\ref{Hgamma}), ${\cal J}(\Delta)$ in (\ref{calJ}), and ${\cal S}(\Delta)$ in 
(\ref{calS}) suggest that the ``natural'' choices for the three scales are
$\mu_h=m_b$, $\mu_i=\sqrt{m_b\Delta}$, and $\mu_0=\Delta$, as this removes all 
logarithms from these expressions. The latter two assignments are supported by
the observation that, for a typical value $\eta\approx 0.25$, the coefficient 
function ${\cal J}(\Delta)$ vanishes near $\mu_i=1.08\sqrt{m_b\Delta}$, while 
$|{\cal S}(\Delta)|$ is minimized near $\mu_0=1.16\Delta$. Below, we will 
adopt the ``natural'' choices as our default values. In practice, a residual 
scale dependence arises because of the truncation of the perturbative 
expansion. Varying the three matching scales about their default values 
provides some information about unknown higher-order perturbative terms.
\item
In the limit where the intermediate and low matching scales $\mu_i$ and 
$\mu_0$ are set equal to the hard matching scale $\mu_h$, our result reduces 
to the conventional formula used in previous analyses of the $B\to X_s\gamma$ 
decay rate. However, this choice cannot be justified on physical grounds.
\item
After RG resummation the decay rate has a non-trivial dependence on the 
photon-energy cut $E_0$ already at leading order in RG-improved perturbation 
theory and at leading power in $\Delta/m_b$, as reflected by the appearance of 
$\Delta^\eta$ in (\ref{Sexpand}). 
\item
In (\ref{rate2}) we have accomplished a complete resummation of 
(parametrically) large logarithms at next-to-next-to-leading logarithmic order 
in RG-improved perturbation theory, which is necessary in order to calculate 
the decay rate with ${\cal O}(\alpha_s)$ accuracy. This is highly non-trivial 
in cases where Sudakov double logarithms are present. Specifically, it means 
that terms of the form $\alpha_s^n L^k$ with $k=(n-1),\dots,2n$ and 
$L=\ln(m_b/\Delta)$ are correctly resummed to all orders in perturbation 
theory. At a given order $\alpha_s^n$, there are $(n+2)$ such terms. To the 
best of our knowledge, a complete resummation at next-to-next-to-leading order 
has never been achieved before. For ease of comparison with the results of 
other authors, we provide in Appendix~A.2 an expansion of our result to second 
order in $\alpha_s$, deriving the coefficients of the terms $\alpha_s^2 L^k$ 
with $k=1,2,3,4$. 
\item
Finally, we stress that the various next-to-leading order corrections in the 
expression for the decay rate obtained from (\ref{rate2}) and (\ref{Sexpand}) 
should be consistently expanded to order $\alpha_s$ before applying our 
results to phenomenology. Such next-to-leading order terms are contained in 
the functions $C_{7\gamma}^{\rm eff}(\mu_h)$, $\overline{m}_b(\mu_h)$, 
$H_\gamma(\mu_h)$, $U_1(\mu_h,\mu_i)$, $U_2(\mu_i,\mu_0)$, $\eta$, and 
$I(E_0)$. For instance, one should expand
\begin{equation}
   \frac{e^{-\gamma_E\eta}}{\Gamma(1+\eta)}
   = \frac{e^{-\gamma_E\eta_0}}{\Gamma(1+\eta_0)}
   \left[ 1 - \frac{\Gamma_0}{\beta_0}
   \left( \frac{\Gamma_1}{\Gamma_0} - \frac{\beta_1}{\beta_0} \right)
   \frac{\alpha_s(\mu_0) - \alpha_s(\mu_i)}{4\pi}\,h(\eta_0) + \dots \right] ,
\end{equation}
where 
$\eta_0=\frac{\Gamma_0}{\beta_0}\ln\frac{\alpha_s(\mu_0)}{\alpha_s(\mu_i)}$ is 
the leading-order expression for $\eta$ (see Appendix~A.1). In practice, these
expansions are readily automatized.
\end{enumerate}

\section{Elimination of pole-scheme parameters}
\label{sec:scheme}

The expression for the function $I(E_0)$ in (\ref{Sexpand}) has been derived 
under the implicit assumption that the $b$-quark mass $m_b$, the related 
parameter $\Delta=m_b-2E_0$, and the HQET parameter $\lambda_1$ are defined in 
the on-shell scheme. While this is most convenient for performing calculations 
using heavy-quark expansions, it is well known that HQET parameters defined in 
the pole scheme suffer from infra-red renormalon ambiguities 
\cite{Bigi:1994em,Beneke:1994sw,Martinelli:1995zw,Neubert:1996zy}. As a 
result, the perturbative expansion in (\ref{Sexpand}) would not be well 
behaved. It is thus necessary to replace the pole mass $m_b$ and the HQET 
parameter $\lambda_1$ in favor of some physical, short-distance parameters. 

For our purposes, the ``shape-function scheme'' defined in \cite{Bosch:2004th} 
provides for a particularly suitable definition of the heavy-quark mass and 
kinetic energy. A look at (\ref{calIdef}) shows that the pole mass actually 
never enters the expression for the decay rate. Rather, a factor 
$(M_B-\hat\omega)^3$ appears under the integral over the shape function, which 
can be traced back to the factor $(M_B-P_+)^3=8E_\gamma^3$ in the original 
expression for the rate in (\ref{ff}). Roughly speaking, it is the average 
value of $\hat\omega$ that determines the value of the difference 
$(M_B-\hat\omega)$. This observation is the basis of the shape-function 
scheme. The idea is that a good estimate of the right-hand side of 
(\ref{calIdef}) can be obtained using the mean-value theorem, i.e., by 
replacing $\hat\omega$ with its mean value defined as
\begin{equation}
   \langle\hat\omega\rangle_\Delta =
   \frac{\int_0^{\Delta_E}\!d\hat\omega\,\hat\omega\,\hat S(\hat\omega,\mu_0)}
        {\int_0^{\Delta_E}\!d\hat\omega\,\hat S(\hat\omega,\mu_0)}
   \equiv \bar\Lambda(\Delta,\mu_0) = M_B - m_b(\Delta,\mu_0) \,.
\end{equation}
Here $m_b(\Delta,\mu_0)$ is the running shape-function mass defined in 
\cite{Bosch:2004th}, which depends on a hard cutoff $\Delta$ in addition to 
the renormalization scale $\mu_0$. The quantity $\Delta$ in the shape-function 
scheme is defined by the implicit equation
\begin{equation}
   \Delta = \Delta_E - \bar\Lambda(\Delta,\mu_0)
   = m_b(\Delta,\mu_0) - 2E_0 \,.
\end{equation}
(For simplicity, we write $\Delta$ instead of the more correct notation 
$\Delta(\Delta,\mu_0)$.) Likewise, we define a kinetic-energy parameter 
$\mu_\pi^2(\Delta,\mu_0)$ via the variance of $\hat\omega$,
\begin{equation}
   \langle\hat\omega^2\rangle_\Delta - \langle\hat\omega\rangle_\Delta^2
   = \frac{\int_0^{\Delta_E}\!d\hat\omega\,\hat\omega^2\,
           \hat S(\hat\omega,\mu_0)}
          {\int_0^{\Delta_E}\!d\hat\omega\,\hat S(\hat\omega,\mu_0)}
   - \langle\hat\omega\rangle_\Delta^2
   \equiv \frac{\mu_\pi^2(\Delta,\mu_0)}{3} \,.
\end{equation}
The shape-function scheme provides a physical, short-distance definition of 
$m_b$ and $\mu_\pi^2$, which can be related to any other short-distance 
definition of these parameters using perturbation theory. The explicit form of 
these relations for some common renormalization schemes can be found in 
\cite{Bosch:2004th}. Here we need the relations to the parameters defined in 
the pole scheme. They are
\begin{eqnarray}\label{RSchange1}
   m_b^{\rm pole} 
   &=& m_b(\Delta,\mu_0) + \Delta\,\frac{C_F\alpha_s(\mu_0)}{\pi}
    \left[ \left( 1 - 2\ln\frac{\Delta}{\mu_0} \right) 
    + \frac23\,\frac{\mu_\pi^2(\Delta,\mu_0)}{\Delta^2}\,
    \ln\frac{\Delta}{\mu_0} \right] + \dots \,, \nonumber\\
   - \lambda_1
   &=& \mu_\pi^2(\Delta,\mu_0) \left[ 1 + \frac{C_F\alpha_s(\mu_0)}{\pi}
    \left( - 3\ln\frac{\Delta}{\mu_0} - \frac12 \right) \right]
    + 3\Delta^2\,\frac{C_F\alpha_s(\mu_0)}{\pi}\,\ln\frac{\Delta}{\mu_0}
    + \dots \,.
\end{eqnarray}
The corresponding relation for $\Delta^{\rm pole}$ follows from the fact that 
$\Delta^{\rm pole}=m_b^{\rm pole}-2E_0$. In order to introduce the parameters 
defined in the shape-function scheme, we perform these replacements in the 
expression for $I(E_0)$ in (\ref{Sexpand}) and expand the result consistently 
to order $\alpha_s$. (In the next-to-leading order terms we can simply replace 
the parameters of the pole scheme by the corresponding parameters of the 
shape-function scheme.) 

While the above choice appears most natural to us, it is by no means unique.
For instance, we may avoid using the running quantities $m_b(\mu_f,\mu)$ and
$\mu_\pi^2(\mu_f,\mu)$ with ``off-diagonal'' scale choices $\mu_f\ne\mu$ by 
using instead the parameters $m_b(\mu,\mu)$ and $\mu_\pi^2(\mu,\mu)$, which 
are related to the parameters in the pole scheme by the simpler relations 
\begin{equation}\label{RSchange2}
   m_b^{\rm pole} = m_b(\mu,\mu) + \mu\,\frac{C_F\alpha_s(\mu)}{\pi}
    + \dots \,,\qquad
   - \lambda_1 = \mu_\pi^2(\mu,\mu) \left[ 1
    - \frac{C_F\alpha_s(\mu)}{2\pi} \right] + \dots \,.
\end{equation}
The parameter $\Delta$ is now determined by the equation 
$\Delta=m_b(\mu,\mu)-2E_0$. The scale $\mu$ could naturally be taken to be 
$\mu_0$. Alternatively, we may use the parameters of the shape-function scheme 
defined at a fixed reference scale $\mu_*=1.5$\,GeV, at which their values 
have been determined to be $m_b(\mu_*,\mu_*)=(4.65\pm 0.07)$\,GeV and 
$\mu_\pi^2(\mu_*,\mu_*)=(0.27\pm 0.07)\,\mbox{GeV}^2$ \cite{Bosch:2004th}. 
These determinations are based on various sources of phenomenological 
information, including $\Upsilon$ spectroscopy and moments of inclusive 
$B$-meson decay spectra. In our numerical analysis in 
Section~\ref{sec:numerics} we will present results for different variants of 
the shape-function scheme.

\section{Kinematic power corrections}
\label{sec:power}

The results of the previous section provide a complete description of the 
$B\to X_s\gamma$ decay rate at leading order in the $1/m_b$ expansion, where 
the two-step matching procedure QCD\,$\to$\,SCET\,$\to$\,HQET is well 
understood. The matching coefficients and anomalous dimensions are known to 
the required order, so that the scale separation and RG resummation can be 
carried out with next-to-next-to-leading logarithmic accuracy. For practical 
applications, however, it is necessary to also include corrections arising at 
higher orders in the heavy-quark expansion. The leading non-perturbative 
corrections proportional to the HQET parameter $\lambda_1$ (or $\mu_\pi^2$) 
have already been included above. More important, however, are ``kinematic'' 
power corrections of order $(\Delta/m_b)^n$, which are not associated with new 
hadronic parameters. Unlike the non-perturbative corrections, these effects 
arise already at first order in $\Delta/m_b$, and they are numerically 
dominant in the region where $\Delta\gg\Lambda_{\rm QCD}$. Technically, the 
kinematic power corrections arise in the matching of QCD correlators onto 
higher-dimensional SCET and HQET operators.

The corresponding terms are known in fixed-order perturbation theory, without 
scale separation and RG resummation \cite{Greub:1996tg,Ali:1995bi} (see also 
\cite{Kagan:1998ym}). To perform a complete RG analysis of even the 
first-order terms in $\Delta/m_b$ is beyond the scope of the present work. 
Since for typical values of $E_0$ the power corrections only account for about 
15\% of the $B\to X_s\gamma$ decay rate, an approximate treatment will 
suffice. To motivate it, we note the following two facts: First, while the 
anomalous dimensions of the relevant subleading SCET and HQET operators are 
only known for a few cases \cite{Hill:2004if}, the leading Sudakov double 
logarithms are determined by the cusp anomalous dimension and thus are the 
same as for the terms of leading power. The reason is that the cusp anomalous 
dimension has a geometric origin. In the present case, it results from a 
product of time-like and light-like Wilson lines describing heavy and 
hard-collinear quark fields, respectively \cite{Becher:2003kh}. The leading 
Sudakov double logarithms are therefore the same as those resummed into the 
exponents $S(\mu_h,\mu_i)$ and $S(\mu_0,\mu_i)$ contained in the evolution 
functions $U_1$ and $U_2$ in (\ref{U1}) and (\ref{U2}). Secondly, all 
power-suppressed terms of order $(\Delta/m_b)^n$ are associated with gluon 
emission into the hadronic final state $X_s$. Because of the kinematic 
restriction to low-mass final states, i.e.\ $M_X^2\le M_B\Delta_E$, the 
emitted gluon can only be hard-collinear or soft, but not hard. One should 
therefore associate a coupling $\alpha_s(\mu_i)$ or $\alpha_s(\mu_0)$ with 
these terms. The leading power corrections are then 
of order $\alpha_s(\mu)\,\delta\ln\delta$ with $\delta=\Delta/m_b$ and 
$\mu\sim\mu_i$ or $\mu_0$. After RG resummation, they can give rise to effects 
of order $\eta\,\delta$, which are formally of zeroth order in the coupling 
constant. Not resolving the scale ambiguity for such terms introduces an 
uncertainty that is at most of order $\alpha_s^2\ln^2\delta$.

In order to at least partially account for resummation effects, we proceed as
follows: We include the known power corrections from real gluon emission and 
associate the coupling $\alpha_s(\mu_i)$ with them. The Wilson coefficients 
$C_i$ are evaluated at the hard scale $\mu_h$. We then multiply the answer 
with the evolution function $U_1$. This accounts correctly for the leading 
Sudakov logarithms in the evolution from the hard scale $\mu_h$ to the 
intermediate scale $\mu_i$. While the conventional parton-model calculation of 
the $B\to X_s\gamma$ decay rate is performed with on-shell $b$ quarks, we add 
a residual momentum such that $p_b=m_b v+k$. In the light-cone component 
$n\cdot p_b=m_b+n\cdot k$ we keep the $n\cdot k$ piece, because it is of the 
same order as the corresponding component $n\cdot p_{hc}$ of a hard-collinear 
momentum. In all other components we neglect $k$. This accounts for some, but 
not all shape-function effects. The net result is that we must replace 
$m_b\to M_B-\hat\omega$ (and hence $\Delta_E\to\Delta_E-\hat\omega$) in the 
parton-model calculation, and then convolute the result with the leading-order 
shape function. In the approximation where the small parameter 
$\varepsilon_{\rm CKM}$ in (\ref{epsCKM}) is set to zero (which is an 
excellent approximation given that we are dealing with power-suppressed 
effects), this yields
\begin{eqnarray}\label{step1}
   \Gamma_{\bar B\to X_s\gamma}^{\rm power}(E_0)
   &=& \frac{G_F^2\alpha}{32\pi^4}\,|V_{tb} V_{ts}^*|^2\,
    \overline{m}_b^2(\mu_h)\,U_1(\mu_h,\mu_i)
    \int_0^{\Delta_E}\!d\hat\omega\,\hat S(\hat\omega,\mu_i)\,
    (M_B-\hat\omega)^3 \\
   &&\hspace{-2.0cm}\times \Bigg[ \frac{C_F\alpha_s(\mu_i)}{4\pi}\!\!
    \sum_{\scriptsize\begin{array}{c} i,j=1,7,8 \\[-0.1cm] i\le j \end{array}}
    \!\!\mbox{Re}\,\Big(C_i^*(\mu_h)\,C_j(\mu_h)\Big)\,
    \hat f_{ij}\Big( \frac{\Delta_E-\hat\omega}{M_B-\hat\omega} \Big)
    - \mbox{Re}\,\Big(C_1^*(\mu_h)\,C_{7\gamma}^{\rm eff}(\mu_h)\Big)\,
    \frac{\lambda_2}{9m_c^2} \Bigg] \,. \nonumber 
\end{eqnarray}
The functions $\hat f_{ij}(\delta)$ vanish linearly with $\delta$ and so are 
of order $\Delta/m_b$. They coincide with $3f_{ij}(\delta)$ in 
\cite{Kagan:1998ym} except for the case of $\hat f_{77}(\delta)$, which 
requires an additional subtraction due to the fact that the function 
$p_3(\eta,\delta)$ in (\ref{Sexpand}) already contains some power corrections 
resulting from the presence of the factor $(M_B-P_+)^3$ in (\ref{ff}). We find
\begin{equation}
   \hat f_{77}(\delta) = 3f_{77}(\delta)
   - \delta\,(12\ln\delta + 9) + \delta^2 \left( 6\ln\delta + \frac{15}{2}
   \right) - \delta^3 \left( \frac43\ln\delta + \frac{17}{9} \right) .
\end{equation}
The relevant expressions are
\begin{eqnarray}\label{fijfunctions}
   \hat f_{77}(\delta) &=& \delta + \frac{17\delta^2}{2}
    - \frac{23\delta^3}{9}
    - \delta \left( 16 - 7\delta + \frac{4\delta^2}{3} \right) \ln\delta \,,
    \nonumber\\
   \hat f_{88}(\delta) &=& \frac49\,\Bigg\{ L_2(1-\delta)
    - \frac{\pi^2}{6} + 2\ln(1-\delta) - \frac{\delta}{4}\,(2+\delta)
    \ln\delta \nonumber\\
   &&\mbox{}+ \frac{7\delta}{4} + \frac{3\delta^2}{4} - \frac{\delta^3}{6}
    - \left[ \delta + \frac{\delta^2}{2} + 2\ln(1-\delta) \right]
    \ln\frac{m_b}{m_s} \Bigg\} \,, \nonumber\\
   \hat f_{78}(\delta) &=& \frac83 \left[ L_2(1-\delta)
    - \frac{\pi^2}{6} - \delta\ln\delta + \frac{9\delta}{4}
    - \frac{\delta^2}{4} + \frac{\delta^3}{12} \right] , \nonumber\\
   \hat f_{11}(\delta) &=& \frac{16}{9} \int_0^1\!dx\,
    (1-x)(1-x_\delta)\,\left|\,\frac{z}{x}\,
    G\!\left(\frac{x}{z}\right) + \frac12\,\right|^2 , \nonumber\\
   \hat f_{17}(\delta) &=& -3\hat f_{18}(\delta)
    = - \frac83 \int_0^1\!dx\,x(1-x_\delta)\,
    \mbox{Re} \left[\, \frac{z}{x}\,G\!\left(\frac{x}{z}\right) + \frac12
    \,\right] ,
\end{eqnarray}
where $x_\delta=\mbox{max}(x,1-\delta)$, as previously $z=(m_c/m_b)^2$, and
\begin{equation}\label{Gdef}
   G(t) = \left\{ \begin{array}{ll}
    -2\arctan^2\!\sqrt{t/(4-t)} & ~;~ t<4 \,, \\[0.1cm]
    2 \left( \ln\!\Big[(\sqrt{t}+\sqrt{t-4})/2\Big]
    - \displaystyle\frac{i\pi}{2} \right)^2 & ~;~ t\ge 4 \,.
   \end{array} \right.
\end{equation}
The next step is to account for the evolution between $\mu_i$ and $\mu_0$, and 
to evaluate the shape-function integrals for $\Delta\gg\Lambda_{\rm QCD}$ 
using the techniques described in Section~\ref{sec:newOPE}. From 
(\ref{sonice}) and (\ref{Sparton}), we find 
\begin{equation}
   \hat S_{\rm parton}(\hat\omega,\mu_i)
   = U_2(\mu_i,\mu_0)\,\frac{e^{-\gamma_E\eta}}{\Gamma(\eta)}\,
   \frac{\theta(\hat\omega-\bar\Lambda)}
        {\mu_0^\eta(\hat\omega-\bar\Lambda)^{1-\eta}}
   + \dots \,,
\end{equation}
where $\bar\Lambda$ is defined in the shape-function scheme (see 
Section~\ref{sec:scheme}), and the dots represent terms of order 
$\alpha_s(\mu_0)$ and higher-order non-perturbative corrections, which we 
consistently neglect. Inserting this result into (\ref{step1}) yields
\begin{eqnarray}\label{dGamma}
   \Gamma_{\bar B\to X_s\gamma}^{\rm power}(E_0)
   &=& \frac{G_F^2\alpha}{32\pi^4}\,|V_{tb} V_{ts}^*|^2\,
    \overline{m}_b^2(\mu_h)\,U_1(\mu_h,\mu_i)\,U_2(\mu_i,\mu_0)\,
    \frac{e^{-\gamma_E\eta}}{\Gamma(1+\eta)}\, 
     m_b^3 \left( \frac{\Delta}{\mu_0} \right)^\eta
     p_3\Big(\eta,\frac{\Delta}{m_b}\Big) \nonumber \\
   &&\hspace{-2.0cm}\times \Bigg[ \frac{C_F\alpha_s(\mu_i)}{4\pi}\!\!
    \sum_{\scriptsize\begin{array}{c} i,j=1,7,8 \\[-0.1cm] i\le j \end{array}}
    \!\!\mbox{Re}\,\Big(C_i^*(\mu_h)\,C_j(\mu_h)\Big)\,
    F_{ij}\Big(\eta,\frac{\Delta}{m_b}\Big)
    - \mbox{Re}\,\Big(C_1^*(\mu_h)\,C_{7\gamma}^{\rm eff}(\mu_h)\Big)\,
    \frac{\lambda_2}{9m_c^2} \Bigg] \,, \qquad
\end{eqnarray}
where
\begin{equation}
   F_{ij}(\eta,\delta)
   = \frac{1}{p_3(\eta,\delta)}\,\int_0^1\!dy\,\eta\,y^{\eta-1}
   (1-y\delta)^3\,\hat f_{ij}\!\left( \frac{(1-y)\delta}{1-y\delta} \right) .
\end{equation}
The definition of the ``smeared'' functions $F_{ij}(\eta,\delta)$ is such that 
$F_{ij}(0,\delta)=\hat f_{ij}(\delta)$, $F_{ij}(\eta,0)=\hat f_{ij}(0)=0$, and 
$F_{ij}(\eta,1)=\hat f_{ij}(1)$. 

The result (\ref{dGamma}) has the desired features that the leading Sudakov
double logarithms are correctly resummed in the product $U_1\,U_2$, and that
the gluon-emission terms are associated with a low-scale coupling constant 
that is larger than $\alpha_s(\mu_h)$. However, we stress that while the 
result is correct when expanded in fixed-order perturbation theory to first 
order in $\alpha_s$, the resummation of single logarithmic terms is only 
approximate. After a complete RG resummation, terms of the form 
$\alpha_s\ln(\Delta/m_b)$, which arise from the $\ln\delta$ terms in the 
expressions for the functions $\hat f_{ij}$, would be resummed into functions 
of $\eta$, e.g.\
\begin{equation}\label{simplerepl}
   \frac{C_F\alpha_s(\mu_i)}{\pi}\,\ln\frac{m_b}{\Delta}
   \to \eta + \frac{C_F\alpha_s(\mu_i)}{\pi}\,\ln\frac{m_b\Delta}{\mu_i^2}
   - \frac{2C_F\alpha_s(\mu_0)}{\pi}\,\ln\frac{\Delta}{\mu_0} + \dots \,.
\end{equation}
The correct answer will contain more complicated functions of $\eta$ as well 
as non-logarithmic next-to-leading-order corrections at the scales $\mu_i$ and 
$\mu_0$. 

While we expect that (\ref{dGamma}) gives a good approximation for the 
power-suppressed contributions to the $B\to X_s\gamma$ decay rate, it would be 
important and conceptually interesting to explore the structure of power 
corrections further, using the effective field-theory technology developed 
here and in \cite{Bosch:2004th}. It should be possible (with a significant 
amount of work) to resolve the scale ambiguity for the first-order power 
corrections in $\Delta/m_b$. Also, an effective field-theory analysis would 
allow a more rigorous description of certain non-perturbative effects, such as 
the $\lambda_2/m_c^2$ term in (\ref{step1}), which models a long-distance 
contribution related to charm-penguin diagrams 
\cite{Voloshin:1996gw,Ligeti:1997tc}, or the logarithmic mass singularity 
regularized by $m_s$ in the expression for $\hat f_{88}$ in 
(\ref{fijfunctions}), which is related to fragmentation effects 
\cite{Kapustin:1995fk}. More generally, such an analysis would provide a 
transparent power counting for any long-distance contributions involving soft 
partons (not only heavy quarks) in the MSOPE.

\section{Ratios of decay rates}
\label{sec:ratios}

The contributions from the three different short-distance scales entering our 
central result (\ref{rate2}) and the associated theoretical uncertainties can 
be disentangled by taking ratios of decay rates. Some ratios probe truly
short-distance physics (i.e., physics above the scale $\mu_h\sim m_b$) and so 
remain unaffected by the new theoretical results obtained in this paper. For 
some other ratios, the short-distance physics associated with the hard scale
cancels to a large extent, so that one probes physics at the intermediate and 
low scales, irrespective of the short-distance structure of the theory. These 
ratios are important, because they are insensitive to New Physics and just 
probe the interplay of hard-collinear and low scales in the process. Below, we 
investigate examples of both classes of ratios.

\subsection{Ratios insensitive to low-scale physics}

Most importantly, physics beyond the Standard Model may affect the theoretical 
results for the $B\to X_s\gamma$ branching ratio and CP asymmetry only via the 
Wilson coefficients of the various operators in the effective weak 
Hamiltonian. (An exception are unconventional New Physics scenarios with new 
light particles, such as a supersymmetric model with light gluinos and 
$\tilde b$ squarks considered in \cite{Becher:2002ue}.) As a result, the ratio 
of the $B\to X_s\gamma$ decay rate in a New-Physics model relative to that in 
the Standard Model remains largely unaffected by the resummation effects 
studied in the present work. From (\ref{rate2}), we obtain
\begin{equation}
   \frac{\Gamma_{\bar B\to X_s\gamma}|_{\rm NP}}
        {\Gamma_{\bar B\to X_s\gamma}|_{\rm SM}}
   = \frac{|H_\gamma(\mu_h)|_{\rm NP}^2}{|H_\gamma(\mu_h)|_{\rm SM}^2}
   + \mbox{power corrections.}
\end{equation}
The power corrections would introduce some mild dependence on the intermediate 
and low scales $\mu_i$ and $\mu_0$, as well as on the cutoff $E_0$.

Another important example is the direct CP asymmetry in $B\to X_s\gamma$ 
decays, for which we obtain
\begin{equation}
   A_{\rm CP}
   = \frac{\Gamma_{\bar B\to X_s\gamma} - \Gamma_{B\to X_{\bar s}\gamma}}
          {\Gamma_{\bar B\to X_s\gamma} + \Gamma_{B\to X_{\bar s}\gamma}}
   = \frac{|H_\gamma(\mu_h)|^2 - |\overline{H}_\gamma(\mu_h)|^2}
          {|H_\gamma(\mu_h)|^2 + |\overline{H}_\gamma(\mu_h)|^2}
   + \mbox{power corrections,}
\end{equation}
where $\overline{H}_\gamma(\mu_h)$ is obtained by CP conjugation, which in the 
Standard Model amounts to replacing
$\varepsilon_{\rm CKM}\to\varepsilon_{\rm CKM}^*$ in (\ref{Hgamma}). It 
follows that the predictions for the CP asymmetry in the Standard Model and 
various New Physics scenarios presented in \cite{Kagan:1998bh} remain largely  
unaffected by our considerations.

\subsection{Ratios sensitive to low-scale physics}
\label{sec:lowratios}

The multi-scale effects studied in this work result from the fact that in 
practice the $B\to X_s\gamma$ decay rate is measured with a restrictive cut on 
the photon energy. As we have pointed out, this introduces sensitivity to the 
scales $\mu_i\sim\sqrt{m_b\Delta}$ and $\mu_0\sim\Delta=m_b-2E_0$ in addition
to the hard scale $\mu_h\sim m_b$. These complications would be absent if it
were possible to measure the fully inclusive rate. It is 
convenient to define a function $F(E_0)$ as the ratio of the $B\to X_s\gamma$ 
decay rate with a cut $E_0$ divided by the total rate, 
\begin{equation}\label{Fdef}
   F(E_0) = \frac{\Gamma_{\bar B\to X_s\gamma}(E_0)}
                 {\Gamma_{\bar B\to X_s\gamma}(E_*)} \,. 
\end{equation}
Because of a logarithmic soft-photon divergence for very low energy, it is 
conventional to define the ``total'' inclusive rate as the rate with a very 
low cutoff $E_*=m_b/20$ \cite{Kagan:1998ym}. The denominator in the expression 
for $F(E_0)$ can be evaluated using a conventional OPE, which corresponds to 
setting all three matching scales equal to $\mu_h$. The numerator is given by 
our expression in (\ref{rate2}), supplemented by the power corrections in 
(\ref{dGamma}). We obtain
\begin{eqnarray}\label{Fmagic}
   F(E_0)
   &=& U_1(\mu_h,\mu_i)\,U_2(\mu_i,\mu_0)\,
    \frac{e^{-\gamma_E\eta}}{\Gamma(1+\eta)}
    \left( \frac{\Delta}{\mu_0} \right)^\eta \\
   &&\hspace{-1.0cm}\times
    \frac{\displaystyle {\cal D}(\Delta) \left[ p_3(\eta,\delta)
      + \frac{\eta(\eta-1)}{2}\,\frac{(-\lambda_1)}{3\Delta^2} \right]
      + p_3(\eta,\delta)\,\frac{C_F\alpha_s(\mu_i)}{4\pi}
      \sum_{i\le j}\,\mbox{Re}\,
      \frac{C_i^*(\mu_h)\,C_j(\mu_h)}{|C_{7\gamma}^{\rm eff}(\mu_h)|^2}\,
      F_{ij}(\eta,\delta)}
     {\displaystyle 1 + \frac{C_F\alpha_s(\mu_h)}{4\pi}\,\Bigg[ 
      {\cal H}(\delta_*) + \sum_{i\le j}\,\mbox{Re}\,
      \frac{C_i^*(\mu_h)\,C_j(\mu_h)}{|C_{7\gamma}^{\rm eff}(\mu_h)|^2}\,
      \hat f_{ij}(\delta_*) \Bigg]} \,, \nonumber
\end{eqnarray}
where $\delta=\Delta/m_b$, $\delta_*=1-2E_*/m_b=0.9$, and
\begin{eqnarray}
   {\cal D}(\Delta)
   &=& 1 + \frac{C_F\alpha_s(\mu_i)}{4\pi}\,{\cal J}(\Delta)
    + \frac{C_F\alpha_s(\mu_0)}{4\pi}\,{\cal S}(\Delta) \,, \nonumber\\
   {\cal H}(\delta_*)
   &=& 4\ln^2\frac{m_b}{\mu_h} - 10\ln\frac{m_b}{\mu_h}
    - 2\ln^2\delta_* - 7\ln\delta_* + 7 - \frac{7\pi^2}{6}
    \nonumber\\
   &&\mbox{}+ \delta_*\,(12\ln\delta_* + 9) 
    - \delta_*^2 \left( 6\ln\delta_* + \frac{15}{2} \right)
    + \delta_*^3 \left( \frac43\ln\delta_* + \frac{17}{9}
    \right) \nonumber\\
   &&\mbox{}+ \left( 2\ln\delta_* + \frac32 \right)
    \frac{(-\lambda_1)}{3\delta_*^2 m_b^2} \,.
\end{eqnarray} 
The result (\ref{Fmagic}) is RG invariant and so (formally) independent of the
three matching scales $\mu_h$, $\mu_i$, and $\mu_0$, and at leading power 
it is insensitive to the hard matching corrections contained 
in $H_\gamma(\mu_h)$. To an excellent approximation, the fraction function 
$F(E_0)$ therefore applies to the Standard Model as well as to any New Physics 
scenario. Note also that the $b$-quark mass enters the expression for the 
fraction function only at the level of power corrections. The prefactor 
$m_b^3\,\overline{m}_b^2(\mu_h)$, which multiplies the total decay rate, 
cancels out in the ratio (\ref{Fdef}). Finally, we stress that the expression
for $F(E_0)$ given above still refers to the pole scheme. It is necessary to 
eliminate the pole-scheme parameters $m_b$ and $\lambda_1$ in favor of 
physical parameters before using this result.

Another important example of a ratio that is largely insensitive to the hard 
matching contributions is the average photon energy defined as
\begin{equation}\label{Eavgdef}
   \langle E_\gamma\rangle
   = \frac{\displaystyle\int_{E_0}^{M_B/2}\!dE_\gamma\,E_\gamma\,
           \frac{d\Gamma}{dE_\gamma}} 
          {\displaystyle\int_{E_0}^{M_B/2}\!dE_\gamma\,
           \frac{d\Gamma}{dE_\gamma}} \,,
\end{equation}
which has been proposed as a good way to measure the $b$-quark mass or, 
equivalently, the HQET parameter $\bar\Lambda$ 
\cite{Kapustin:1995nr,Ligeti:1999ea}. The impact of shape-function effects on 
the theoretical prediction for this ratio has been studied in 
\cite{Kagan:1998ym,Bigi:2002qq} and was found to be significant. Here we study 
the average photon energy in the MSOPE region, where a model-independent 
prediction can be obtained. It is structurally different from the one obtained 
using the conventional OPE in the sense that contributions associated with 
different scales are disentangled from each other. We find (with 
$\delta=\Delta/m_b$)
\begin{equation}\label{Egamavg}
   \langle E_\gamma\rangle
   = \frac{m_b}{2} \left( 1 - \frac{\lambda_1+3\lambda_2}{2m_b^2} \right)
   \Bigg\{
   \frac{{\cal D}(\Delta)\,p_4(\eta,\delta)}
        {{\cal D}(\Delta)\,p_3(\eta,\delta)}
   + \frac{C_F\alpha_s(\mu_i)}{4\pi}\!\!
   \sum_{\scriptsize\begin{array}{c} i,j=1,7,8 \\[-0.1cm] i\le j
   \end{array}}\!\!\mbox{Re}\,
   \frac{C_i^*(\mu_h) C_j(\mu_h)}{|C_{7\gamma}^{\rm eff}(\mu_h)|^2}\,
   d_{ij}(\delta) \Bigg\} \,,
\end{equation}
where $p_4(\eta,\delta)$ is defined in (\ref{fdef}), and
\begin{equation}
   d_{ij}(\delta)
   = \int_0^\delta\!dx\,\hat f_{ij}(x) - \delta\,\hat f_{ij}(\delta) \,. 
\end{equation}
Analytical expressions for the functions $d_{ij}(\delta)$ are given in 
Appendix~A.3. They vanish quadratically for $\delta\to 0$ and so give very 
small contributions for realistic values of the cutoff. We therefore do not 
include RG resummation effects for these terms. The non-perturbative 
corrections involving the parameters $\lambda_1$ and $\lambda_2$ are taken 
from \cite{Falk:1993dh}. Note that the expression in brackets is a purely 
perturbative result free of hadronic parameters. When expanded in 
fixed-order perturbation theory, our result (\ref{Egamavg}) reduces to an 
expression first obtained in \cite{Kagan:1998ym}.

We stress that the hard scale $\mu_h\sim m_b$ affects the average photon 
energy only via second-order power corrections. This shows that it is not
appropriate to compute the quantity $\langle E_\gamma\rangle$ using a simple 
heavy-quark expansion at the scale $m_b$, which is however done in the 
conventional OPE approach \cite{Kapustin:1995nr,Ligeti:1999ea}. This 
observation is important, because information about moments of the 
$B\to X_s\gamma$ photon spectrum is sometimes used in global fits to determine 
the CKM matrix element $|V_{cb}|$ along with HQET parameters. Keeping only the 
leading power corrections, which is a very good approximation, the above 
expression simplifies to
\begin{eqnarray}\label{Eavgapprox}
   \langle E_\gamma\rangle 
   &=& \frac{m_b}{2} - \frac{\Delta}{2(1+\eta)^2} \Bigg[\, \eta(1+\eta)
    + \frac{C_F\alpha_s(\mu_i)}{\pi} \left( \ln\frac{m_b\Delta}{\mu_i^2}
    - h(\eta) - \frac34 - \frac{1}{1+\eta} \right) \nonumber\\
   &&\mbox{}- \frac{2C_F\alpha_s(\mu_0)}{\pi} \left( \ln\frac{\Delta}{\mu_0}
    - h(\eta) + \frac12 - \frac{1}{1+\eta} \right) \Bigg] + \dots \,.
\end{eqnarray}
In this approximation, $\langle E_\gamma\rangle$ only depends on physics at 
the intermediate and low scales $\mu_i$ and $\mu_0$. The next-to-leading order 
perturbative corrections in this formula are numerically quite significant. 
For $E_0=1.8$\,GeV, and taking the default scale choices 
$\mu_i=\sqrt{m_b\Delta}$ and $\mu_0=\Delta$, we find in the pole scheme (using 
$m_b^{\rm pole}=4.8$\,GeV for the purpose of illustration) 
$\langle E_\gamma\rangle\approx%
[2.27+0.29\alpha_s(\sqrt{m_b\Delta})-0.19\alpha_s(\Delta)]\,\mbox{GeV}$. 
Eliminating the pole mass $m_b$ in favor of the $b$-quark mass 
$m_b(\Delta,\Delta)$ defined in the shape-function scheme, we obtain 
$\langle E_\gamma\rangle\approx[2.222+0.254\alpha_s(\sqrt{m_b\Delta})%
+0.009\alpha_s(\Delta)]\,\mbox{GeV}\approx 2.30$\,GeV. When the $b$-quark mass 
is defined in the shape-function scheme, the average photon energy is 
numerically very close to $\frac12 m_b(\Delta,\Delta)\approx 2.33$\,GeV, 
meaning that the first term in (\ref{Eavgapprox}) dominates. Note also that 
the correction proportional to the low-scale coupling $\alpha_s(\Delta)$ is 
largely reduced in this scheme, ensuring an improved perturbative behavior.

\section{Numerical results}
\label{sec:numerics}

\begin{table}[t]
\centerline{\parbox{14cm}{\caption{\label{tab:inputs}
Compilation of input parameters entering the numerical analysis. The top-quark 
mass enters the expressions for the Wilson coefficients $C_i$. The 
strange-quark mass is required as an infra-red regulator in 
(\ref{fijfunctions}). Only the real part of $\varepsilon_{\rm CKM}$ is needed 
for the calculations in this work.}}}
\vspace{0.1cm}
\begin{center}
\begin{tabular}{|c|c|c|}
\hline\hline
Parameter & Value & Source \\
\hline\hline
$m_b(\mu_*,\mu_*)$ [GeV] & $4.65\pm 0.07$ & \cite{Bosch:2004th} \\
$\mu_\pi^2(\mu_*,\mu_*)$ [GeV$^2$] & $0.27\pm 0.07$ & \cite{Bosch:2004th} \\
\hline
$\overline{m}_b(\overline{m}_b)$ [GeV] & $4.25\pm 0.08$
 & \cite{Beneke:1999fe} \\
$\overline{m}_c(\overline{m}_c)$ [GeV] & $1.25\pm 0.15$
 & \cite{Eidelman:2004wy} \\
$m_t^{\rm pole}$ [GeV] & $178.0\pm 4.3$ & \cite{Abazov:2004cs} \\
$m_s/m_b$ & 0.02 & \cite{Kagan:1998ym} \\
\hline
$\tau_B$ [ps] & $1.604\pm 0.016$ & \cite{Eidelman:2004wy} \\
$\alpha_s(M_Z)$ & $0.1187\pm 0.0020$ & \cite{Eidelman:2004wy} \\
\hline
$|V_{ts}^* V_{tb}|$ [$10^{-3}$] & $40.4_{\,-0.6}^{\,+1.4}$
 & \cite{Charles:2004jd} \\
$\mbox{Re}(\varepsilon_{\rm CKM})$ [$10^{-3}$] & $9.8_{\,-4.2}^{\,+5.1}$
 & \cite{Charles:2004jd} \\
\hline
$\lambda_1$ [GeV$^2$] & $-0.25\pm 0.20$
 & \cite{Cronin-Hennessy:2001fk,Aubert:2003dr} \\
$\lambda_2$ [GeV$^2$] & 0.12 & $\frac14\,(M_{B^*}^2-M_B^2)$ \\
\hline\hline
\end{tabular}
\end{center}
\end{table}

We are now ready to present the phenomenological implications of our findings. 
Table~\ref{tab:inputs} contains a list of the input parameters entering the 
analysis together with their present uncertainties. We have inflated the error 
on $\lambda_1$ obtained by averaging the values quoted in 
\cite{Cronin-Hennessy:2001fk,Aubert:2003dr} from 0.06\,GeV$^2$ to 
0.20\,GeV$^2$, taking into account that this parameter is affected by 
infra-red renormalon ambiguities \cite{Martinelli:1995zw,Neubert:1996zy}. We
vary the quark masses $m_c$ and $m_b$ independently, in which case
$\sqrt{z}=m_c(\mu_h)/m_b(\mu_h)=0.221\pm 0.027$. Additional uncertainties
related to the possibility that the proper normalization point for the 
charm-quark mass in penguin loop graphs may be significantly lower than the 
hard scale $\mu_h$ are considered part of the perturbative error.

The most important correlations between input parameters are implemented as 
follows. We consider the two $b$-quark masses $m_b(\mu_*,\mu_*)$ and 
$\overline{m}_b(\overline{m}_b)$ as being fully correlated and vary their 
values simultaneously. The same applies to the values of the parameters 
$\mu_\pi^2(\mu_*,\mu_*)$ and $\lambda_1$. Next, we use that the value of $m_b$ 
is strongly anti-correlated with that of $|V_{cb}|$, because the most precise 
determination of $m_b$ is obtained from the analysis of $B\to X_c\,l\,\nu$ 
decay distributions. A recent study in \cite{Aubert:2004aw} quotes a 
correlation coefficient $c=-0.49$ between $m_b$ and $|V_{cb}|$. CKM unitarity 
ensures that $|V_{cb}|$ is to a very good approximation equal to the product 
$|V_{ts}^* V_{tb}|$, so that the same anti-correlation can be assumed between 
$m_b$ and $|V_{ts}^* V_{tb}|$. 

Before presenting our results, we reiterate that to apply the formulae derived 
in this work we must first eliminate the parameters $m_b$ and $\lambda_1$ 
defined in the pole scheme in terms of physical parameters defined in the 
shape-function scheme and then expand the answer consistently to 
${\cal O}(\alpha_s)$, treating ratios such as 
$\alpha_s(\mu_i)/\alpha_s(\mu_h)$ and $\alpha_s(\mu_0)/\alpha_s(\mu_i)$ as 
${\cal O}(1)$ parameters. This expansion is readily automatized. Throughout, 
we use the 3-loop expression for the running coupling $\alpha_s(\mu)$ defined 
in the $\overline{\rm MS}$ scheme \cite{Eidelman:2004wy}.

\subsection{Partial \boldmath$B\to X_s\gamma$ branching ratio\unboldmath}

We begin by presenting predictions for the CP-averaged $B\to X_s\gamma$ 
branching fraction with a cutoff $E_\gamma\ge E_0$ applied on the photon 
energy measured in the $B$-meson rest frame. Lowering $E_0$ below 2\,GeV is 
challenging experimentally. The first measurement with $E_0=1.8$\,GeV has 
recently been reported by the Belle Collaboration \cite{Koppenburg:2004fz}. 
It yields\footnote{To obtain the first result we had to undo a theoretical 
correction accounting for the effects of the cut $E_\gamma>1.8$\,GeV, which 
had been applied to the experimental data.}
\begin{eqnarray}\label{belle}
   \mbox{Br}(B\to X_s\gamma) \Big|_{E_0=1.8\,{\rm GeV}}
   &=& (3.38\pm 0.30\pm 0.29)\cdot 10^{-4} \,, \nonumber\\
   \langle E_\gamma\rangle \Big|_{E_0=1.8\,{\rm GeV}}
   &=& (2.292\pm 0.026\pm 0.034)\,\mbox{GeV} \,.
\end{eqnarray}
For $E_0=1.8$\,GeV we have $\Delta\approx 1.1$\,GeV, which is sufficiently 
large to apply the formalism developed in the present work. We will also 
present results for $E_0=1.6$\,GeV because this value has been used in some
theoretical studies, although it has not yet been achieved in an experiment. 
(For comparison, the value $E_0=2.0$\,GeV adopted in the CLEO analysis
\cite{Chen:2001fj} implies $\Delta\approx 0.7$\,GeV, which we believe may be 
too low for a short-distance treatment.) 

We first set all input parameters to their default values and study the
dependence of the branching ratio on the three matching scales $\mu_h$, 
$\mu_i$, and $\mu_0$. The sensitivity of our predictions to variations of the 
matching scales provides an estimate of unknown higher-order perturbative 
corrections. We shall study three different version of the shape-function 
scheme for the definition of the $b$-quark mass and the kinetic-energy 
parameter $\mu_\pi^2$, as discussed in Section~\ref{sec:scheme}. In the first 
scheme (called ``RS~1'') we use the parameters $m_b(\Delta,\mu_0)$ and 
$\mu_\pi^2(\Delta,\mu_0)$ defined in (\ref{RSchange1}). In the second scheme 
(``RS~2'') we instead use $m_b(\mu_0,\mu_0)$ and $\mu_\pi^2(\mu_0,\mu_0)$ from 
(\ref{RSchange2}). Finally, in the third scheme (``RS~3'') we employ the 
parameters $m_b(\mu_*,\mu_*)$ and $\mu_\pi^2(\mu_*,\mu_*)$ renormalized at a 
fixed scale $\mu_*=1.5$\,GeV, at which their values have been determined in 
\cite{Bosch:2004th}. In the schemes RS~1 and RS~2, these reference values are 
evolved to other scales using equations derived in \cite{Bosch:2004th}.

The matching scales are independently varied about their default values 
$\mu_h=m_b$, $\mu_i=\sqrt{m_b\Delta}$, and $\mu_0=\Delta$ by multiplying them 
with factors between $2/3$ and $3/2$. Thus, for $m_b=4.7$\,GeV and 
$E_0=1.8$\,GeV, we vary $\mu_h\in[3.13,7.05]$\,GeV, 
$\mu_i\in[1.52,3.41]$\,GeV, and $\mu_0\in[0.73,1.65]$\,GeV, while for 
$E_0=1.6$\,GeV the latter two ranges are replaced by
$\mu_i\in[1.77,3.98]$\,GeV, and $\mu_0\in[1.0,2.25]$\,GeV. Together, this 
covers a conservative range of scales. The resulting variations of the 
branching ratio are shown in Table~\ref{tab:BrScales}. 

\begin{table}[t]
\centerline{\parbox{14cm}{\caption{\label{tab:BrScales}
$B\to X_s\gamma$ branching ratio with estimates of perturbative uncertainties 
obtained by variation of the matching scales, for three variants of the 
shape-function scheme. See text for explanation.}}}
\vspace{0.1cm}
\begin{center}
\begin{tabular}{|c|c|c|cccc|c|c|}
\hline\hline
$E_0$ & Scheme & Br [$10^{-4}$] & $\mu_h$ & $\mu_i$ & $\mu_0$ & Sum
 & Power Cors.\ & Combined \\
\hline\hline
1.8\,GeV & RS 1 
 & 3.37 & ${}_{\,-0.00}^{\,+0.02}$ & ${}_{\,-0.37}^{\,+0.25}$
 & ${}_{\,-0.03}^{\,+0.41}$ & ${}_{\,-0.37}^{\,+0.48}$
 & ${}_{\,-0.07}^{\,+0.12}$ & ${}_{\,-0.38}^{\,+0.49}$ \\
 & RS 2
 & 3.38 & ${}_{\,-0.00}^{\,+0.02}$ & ${}_{\,-0.37}^{\,+0.25}$
 & ${}_{\,-0.18}^{\,+0.15}$ & ${}_{\,-0.41}^{\,+0.29}$ 
 & ${}_{\,-0.07}^{\,+0.12}$ & ${}_{\,-0.42}^{\,+0.31}$ \\
 & RS 3
 & 3.36 & ${}_{\,-0.00}^{\,+0.02}$ & ${}_{\,-0.37}^{\,+0.25}$
 & ${}_{\,-0.18}^{\,+0.18}$ & ${}_{\,-0.41}^{\,+0.30}$ 
 & ${}_{\,-0.07}^{\,+0.12}$ & ${}_{\,-0.42}^{\,+0.32}$ \\
\hline
1.6\,GeV & RS 1 
 & 3.47 & ${}_{\,-0.00}^{\,+0.02}$ & ${}_{\,-0.39}^{\,+0.28}$
 & ${}_{\,-0.01}^{\,+0.14}$ & ${}_{\,-0.39}^{\,+0.31}$ 
 & ${}_{\,-0.05}^{\,+0.10}$ & ${}_{\,-0.39}^{\,+0.33}$ \\
 & RS 2
 & 3.47 & ${}_{\,-0.00}^{\,+0.02}$ & ${}_{\,-0.39}^{\,+0.28}$
 & ${}_{\,-0.14}^{\,+0.13}$ & ${}_{\,-0.41}^{\,+0.31}$ 
 & ${}_{\,-0.05}^{\,+0.10}$ & ${}_{\,-0.41}^{\,+0.33}$ \\
 & RS 3
 & 3.48 & ${}_{\,-0.00}^{\,+0.02}$ & ${}_{\,-0.39}^{\,+0.28}$
 & ${}_{\,-0.13}^{\,+0.18}$ & ${}_{\,-0.41}^{\,+0.33}$ 
 & ${}_{\,-0.05}^{\,+0.10}$ & ${}_{\,-0.41}^{\,+0.34}$ \\
\hline\hline
\end{tabular}
\end{center}
\end{table}

We observe an excellent stability of our predictions with respect to 
variations of the hard matching scale $\mu_h$. In fact, the sensitivity is so 
small that it cannot reasonably be taken as an indication of the size of 
higher-order terms in the expansion in powers of $\alpha_s(\mu_h)$. The 
sensitivity to variations of the intermediate matching scale $\mu_i$ is more 
pronounced. The numbers suggest that terms of order $\alpha_s^2(\mu_i)$ could 
impact the branching ratio at the 10\% level, which appears entirely 
reasonable given that $\alpha_s(\mu_i)\approx 0.3$. The sensitivity to the low 
matching scale $\mu_0$ turns out to be rather small. The coefficient of the 
$\alpha_s(\mu_0)$ term depends on the scheme adopted for the definition of the 
parameters $m_b$ and $\mu_\pi^2$, and it appears that in the three schemes 
considered here this coefficient is numerically small. While it is not 
guaranteed that this feature will persist in higher orders, the observation of 
good stability at the scale $\mu_0$ suggests that the shape-function scheme 
captures the most important low-scale effects and absorbs them into the 
running $b$-quark mass and the parameter $\mu_\pi^2$. The column labeled 
``Sum'' shows the combined uncertainty obtained by adding the three scale 
variations in quadrature. The next column, labeled ``Power Cors.'', gives an 
estimate of the perturbative uncertainty in our treatment of kinematic power 
corrections, as discussed in Section~\ref{sec:power}. It is obtained by 
studying two variants of the expression (\ref{dGamma}), one where we set 
$p_3\to 1$ and $F_{ij}\to\hat f_{ij}$, and one where in addition we neglect 
all anomalous dimension functions except those governed by 
$\Gamma_{\rm cusp}$. In both cases, we obtain expressions that differ from 
(\ref{dGamma}) by terms that are beyond the accuracy of our calculation. The 
resulting changes in the branching ratio are the same in all schemes and range 
between 1.5 and 3.5\%, corresponding to a 10--25\% uncertainty in the size of 
the power-suppressed contributions themselves. Finally, the last column in the 
table shows our estimates for the total perturbative uncertainty in the 
prediction of the branching ratio, which we find to be of order 10\%, 
significantly larger than previous estimates. For example, the authors of 
\cite{Gambino:2001ew,Buras:2002tp} argued in favor of a total perturbative error 
of only 4\%. 

\begin{table}
\centerline{\parbox{14cm}{\caption{\label{tab:BrPars}
$B\to X_s\gamma$ branching ratio with estimates of theoretical uncertainties 
due to input parameter variations as listed in Table~\ref{tab:inputs}. The 
upper (lower) sign refers to increasing (decreasing) a given input 
parameter.}}}
\vspace{0.1cm}
\begin{center}
\begin{tabular}{|c|c|cccccc|c|}
\hline\hline
$E_0$ & Br [$10^{-4}$] & $m_b$ & $m_c$ & $m_t$ & $|V_{ts}^* V_{tb}|$ & $\tau_B$
 & $\alpha_s(M_Z)$ & Combined \\
\hline\hline
1.8\,GeV & 3.38 & ${}_{\,-0.30}^{\,+0.31}$ & $\mp 0.10$ & $\pm 0.04$
 & ${}_{\,-0.10}^{\,+0.24}$ & $\pm 0.03$ & ${}_{\,-0.08}^{\,+0.07}$
 & ${}_{\,-0.30}^{\,+0.32}$ \\
\hline
1.6\,GeV & 3.48 & ${}_{\,-0.28}^{\,+0.30}$ & ${}_{\,+0.10}^{\,-0.11}$
 & $\pm 0.04$ & ${}_{\,-0.10}^{\,+0.24}$ & $\pm 0.03$ & $\pm 0.10$
 & ${}_{\,-0.29}^{\,+0.32}$ \\
\hline\hline
\end{tabular}
\end{center}
\end{table}

The remaining uncertainties in our predictions are due to input parameter 
variations. They are essentially the same in the three renormalization schemes 
and are summarized in Table~\ref{tab:BrPars} for the case of RS~2. The last 
column shows the combined errors, added in quadrature. They are dominated by 
the uncertainties in the $b$-quark mass and in $|V_{ts}|$, whose significant 
anti-correlation ($c=-0.49$) is taken into account in computing the total 
error. Parameter dependences not shown in the table have a negligible effect 
($<1\%$) on the branching ratio. Note that in contrast to previous authors we 
do not divide the theoretical expression for the $B\to X_s\gamma$ decay rate 
by a semileptonic rate, but present an absolute prediction for the branching 
ratio itself. Once the correlation between parameters is properly taken into 
account, normalizing $\Gamma(B\to X_s\gamma)$ to the semileptonic rate 
$\Gamma(B\to X\,l\,\nu)$ does not lead to a significant reduction of the 
theoretical uncertainties.

The above results can be combined into the new Standard Model predictions
\begin{eqnarray}\label{myresult}
   \mbox{Br}(B\to X_s\gamma) \Big|_{E_0=1.8\,{\rm GeV}}
   &=& (3.38_{\,-0.42}^{\,+0.31}\,[\mbox{pert.}]\,{}_{\,-0.30}^{\,+0.32}\,
    [\mbox{pars.}])\times 10^{-4} \,, \nonumber\\
   \mbox{Br}(B\to X_s\gamma) \Big|_{E_0=1.6\,{\rm GeV}}
   &=& (3.47{}_{\,-0.41}^{\,+0.33}\,[\mbox{pert.}]\,{}_{\,-0.29}^{\,+0.32}\,
    [\mbox{pars.}])\times 10^{-4} \,, 
\end{eqnarray}
where we use the mass renormalization scheme RS~2 as our default. 
The first error refers to the perturbative uncertainty and the second 
one to parameter variations. The first value is in excellent agreement with 
the experimental result (\ref{belle}). Comparing the two results, and naively
assuming Gaussian errors, we find that\footnote{We do not use the CLEO data 
\cite{Chen:2001fj} in deriving this bound, because the choice $E_0=2$\,GeV 
does not allow for a model-independent treatment of the effects of the cut.}
\begin{equation}
   \mbox{Br}(B\to X_s\gamma)_{\rm exp} - \mbox{Br}(B\to X_s\gamma)_{\rm SM}
   < 1.3\cdot 10^{-4} \quad \mbox{(95\% CL)} \,.
 \end{equation}
We stress that, mainly as a result of the enlarged theoretical uncertainty but 
also due to the use of more recent data, this bound is much weaker than the 
one derived in \cite{Gambino:2001ew}, where this difference was found to be 
less than $0.5\cdot 10^{-4}$. Consequently, we obtain weaker constraints on 
New Physics parameters. For instance, for the case of the type-II 
two-Higgs-doublet model, we may use the analysis of \cite{Borzumati:1998tg} to 
obtain the bound
\begin{equation}
   m_{H^+} > \mbox{approx.\ 200\,GeV} \quad \mbox{(95\% CL)} \,,
\end{equation}
which is significantly weaker than the constraints $m_{H^+}>500$\,GeV (at 95\% 
CL) and $m_{H^+}>350$\,GeV (at 99\% CL) found in \cite{Gambino:2001ew}. To 
find the precise numerical value for the bound would require a dedicated 
analysis, which is beyond the scope of this paper.

\subsection{Event fraction \boldmath$F(E_0)$\unboldmath}
\label{sec:FE0}

As an alternative way to discuss the effects of imposing the cut on the photon 
energy, we study the fraction function $F(E_0)$ defined in (\ref{Fdef}), which 
up to power corrections is insensitive to the short-distance physics encoded 
in the Wilson coefficients $C_i$. The sensitivity of $F(E_0)$ to scale 
variations is studied in Table~\ref{tab:FfScales}, which is analogous to 
Table~\ref{tab:BrScales} for the branching ratio. We find that the fraction 
function exhibits a stronger sensitivity to the hard scale $\mu_h$ than the 
branching ratio, changing by about 3\% as $\mu_h$ is varied between $2m_b/3$ 
and $3m_b/2$. The sensitivity to variations of the matching scales $\mu_i$ and 
$\mu_0$ follows the same pattern as in the case of the branching ratio, but 
the variations are somewhat smaller in magnitude. Note that there is a 
difference between the function $F(E_0)$ and the branching ratio as far as the 
dependence on $\mu_0$ is concerned, because the factor $m_b^3$ present in 
(\ref{Sexpand}) and (\ref{dGamma}) cancels in the ratio (\ref{Fmagic}). Since 
in the shape-function scheme the pole mass $m_b$ is expanded in a series in 
$\alpha_s(\mu_0)$, this has an effect on the perturbative expansion. Finally, 
the perturbative uncertainties in the calculation of the power-suppressed 
terms are again at the level of a few percent. Our estimate for the combined 
perturbative error is presented in the last column. 

\begin{table}[t]
\centerline{\parbox{14cm}{\caption{\label{tab:FfScales}
$B\to X_s\gamma$ event fraction $F(E_0)$ with estimates of perturbative 
uncertainties obtained by variation of the matching scales, for three variants 
of the shape-function scheme. See text for explanation.}}}
\vspace{0.1cm}
\begin{center}
\begin{tabular}{|c|c|c|cccc|c|c|}
\hline\hline
$E_0$ & Scheme & $F(E_0)$ [\%] & $\mu_h$ & $\mu_i$ & $\mu_0$ & Sum
 & Power Cors.\ & Combined \\
\hline\hline
1.8\,GeV & RS 1 
 & 89.1 & ${}_{\,-2.2}^{\,+2.4}$ & ${}_{\,-5.0}^{\,+1.4}$
 & ${}_{\,-5.4}^{\,+2.4}$ & ${}_{\,-7.7}^{\,+3.6}$
 & ${}_{\,-2.4}^{\,+4.7}$ & ${}_{\,-8.1}^{\,+5.9}$ \\
 & RS 2
 & 89.1 & ${}_{\,-2.3}^{\,+2.4}$ & ${}_{\,-5.0}^{\,+1.5}$
 & ${}_{\,-4.0}^{\,+2.6}$ & ${}_{\,-6.8}^{\,+3.8}$
 & ${}_{\,-2.4}^{\,+4.7}$ & ${}_{\,-7.2}^{\,+6.0}$ \\
 & RS 3
 & 89.2 & ${}_{\,-2.3}^{\,+2.5}$ & ${}_{\,-5.0}^{\,+1.3}$
 & ${}_{\,-3.7}^{\,+2.5}$ & ${}_{\,-6.6}^{\,+3.8}$ 
 & ${}_{\,-2.4}^{\,+4.6}$ & ${}_{\,-7.0}^{\,+6.0}$ \\
\hline
1.6\,GeV & RS 1 
 & 93.1 & ${}_{\,-2.6}^{\,+2.8}$ & ${}_{\,-5.7}^{\,+2.7}$
 & ${}_{\,-2.3}^{\,+2.6}$ & ${}_{\,-6.7}^{\,+4.7}$ 
 & ${}_{\,-1.9}^{\,+3.9}$ & ${}_{\,-7.0}^{\,+6.1}$ \\
 & RS 2
 & 93.1 & ${}_{\,-2.6}^{\,+2.8}$ & ${}_{\,-5.7}^{\,+2.7}$
 & ${}_{\,-2.5}^{\,+2.4}$ & ${}_{\,-6.8}^{\,+4.6}$ 
 & ${}_{\,-1.9}^{\,+3.9}$ & ${}_{\,-7.1}^{\,+6.0}$ \\
 & RS 3
 & 93.1 & ${}_{\,-2.6}^{\,+2.8}$ & ${}_{\,-5.7}^{\,+2.7}$
 & ${}_{\,-2.3}^{\,+2.6}$ & ${}_{\,-6.7}^{\,+4.7}$ 
 & ${}_{\,-1.9}^{\,+3.9}$ & ${}_{\,-7.0}^{\,+6.1}$ \\
\hline\hline
\end{tabular}
\end{center}
\end{table}

\begin{table}
\centerline{\parbox{14cm}{\caption{\label{tab:FfPars}
$B\to X_s\gamma$ event fraction $F(E_0)$ with estimates of theoretical 
uncertainties due to input parameter variations as listed in 
Table~\ref{tab:inputs}. The upper (lower) sign refers to increasing 
(decreasing) a given input parameter.}}}
\vspace{0.1cm}
\begin{center}
\begin{tabular}{|c|c|cccc|c|}
\hline\hline
$E_0$ & $F(E_0)$ [\%] & $m_b$ & $m_c$ & $\alpha_s(M_Z)$ & $\mu_\pi^2$
 & Combined \\
\hline\hline
1.8\,GeV & 89.1 & ${}_{\,-1.0}^{\,+0.8}$ & $\pm 0.5$ & ${}_{\,+0.9}^{\,-1.1}$
 & $\mp 0.3$ & ${}_{\,-1.6}^{\,+1.3}$ \\
\hline
1.6\,GeV & 93.1 & $\pm 0.3$ & $\pm 0.4$ & $\mp 0.5$ & $\mp 0.1$
 & ${}_{\,-0.8}^{\,+0.7}$ \\
\hline\hline
\end{tabular}
\end{center}
\end{table}

In contrast to the $B\to X_s\gamma$ branching ratio, the fraction function 
$F(E_0)$ is independent of several input parameters (i.e., 
$\overline{m}_b(\overline{m}_b)$, $|V_{ts}^* V_{tb}|$, $\tau_B$, 
$\lambda_{1,2}$, and $\varepsilon_{\rm CKM}$), and it shows a very weak 
sensitivity to variations of the remaining parameters. This is illustrated
in Table~\ref{tab:FfPars}, which summarizes the resulting theoretical 
uncertainties for the case of RS~2. The combined errors are of order 1\% and 
thus almost negligible compared with the perturbative uncertainties. 

In summary, we obtain
\begin{eqnarray}\label{F18}
   F(1.8\,{\rm GeV})
   &=& (89_{\,-7}^{\,+6}\,[\mbox{pert.}]\pm 1\,[\mbox{pars.}])\% \,,
    \nonumber\\
   F(1.6\,{\rm GeV})
   &=& (93_{\,-7}^{\,+6}\,[\mbox{pert.}]\pm 1\,[\mbox{pars.}])\% \,.
\end{eqnarray}
This is the first time that these fractions have been computed in a model 
independent way. The result corresponding to $E_0=1.8$\,GeV may be compared 
with the values $(95.8_{\,-2.9}^{\,+1.3})\%$ and $(95\pm 1)\%$ obtained from 
the study of shape-function models in \cite{Kagan:1998ym} and 
\cite{Bigi:2002qq}, respectively. In these studies, perturbative uncertainties 
have been ignored. A calculation in the conventional OPE approach gives a 
similar result, $(95.2_{\,-2.9}^{\,+1.3})\%$ \cite{Gambino:2001ew}, where the 
authors took the error estimate from \cite{Kagan:1998ym}. In the present 
work, we obtain a smaller central value with a larger uncertainty.

As mentioned in the Introduction, the fraction function $F(E_0)$ can be used 
to combine our study of multi-scale effects with other, independent 
calculations of the total $B\to X_s\gamma$ branching ratio, both in the 
Standard Model and in extensions of it. For instance, we may use the result 
$(3.70\pm 0.31)\times 10^{-4}$ for the total branching ratio in the Standard 
Model obtained from \cite{Gambino:2001ew,Buras:2002tp} (where the error 
contains a 4\% perturbative uncertainty) and combine it with (\ref{F18}) to 
find
\begin{eqnarray}\label{new}
   \mbox{Br}(B\to X_s\gamma) \Big|_{E_0=1.8\,{\rm GeV}}
   &=& (3.30_{\,-0.31}^{\,+0.27}\,[\mbox{pert.}]\pm 0.28\,[\mbox{pars.}])
   \times 10^{-4} \,, \nonumber\\
   \mbox{Br}(B\to X_s\gamma) \Big|_{E_0=1.6\,{\rm GeV}}
   &=& (3.44_{\,-0.30}^{\,+0.27}\,[\mbox{pert.}]\pm 0.27\,[\mbox{pars.}])
   \times 10^{-4} \,.
\end{eqnarray}
Compared with (\ref{myresult}), the perturbative uncertainty is reduced by about 
15--25\%. However, only the reduction in the $\mu_0$ dependence can be taken 
seriously, as the $\mu_i$ dependence is formally the same in (\ref{myresult}) and 
(\ref{new}). The insignificant reduction of the parameter uncertainties is partly 
due to the fact that the authors of \cite{Gambino:2001ew,Buras:2002tp} take 
smaller parameter variations than those in Table~\ref{tab:inputs}, namely 
$\pm 100$\,MeV for $m_c$, $\pm 30$\,MeV for $m_b$, and an error on the ratio 
$|V_{ts}^* V_{tb}|/|V_{cb}|$ that is half as big as what one would obtain using 
the global fit results compiled in \cite{Charles:2004jd}. 

As a final remark, we compare our results for the branching ratio with a cut 
at $E_0=1.6$\,GeV in (\ref{myresult}) and (\ref{new}) with the benchmark value 
$(3.57\pm 0.30)\times 10^{-4}$ corresponding to the most recent calculation 
\cite{Buras:2002tp} published prior to the present work. Our central values are 
about 3\% lower and, more importantly, the theoretical uncertainties we find are 
about 50\% larger.

\subsection{Average photon energy}

\begin{table}[t]
\centerline{\parbox{14cm}{\caption{\label{tab:EavgScales}
Scale dependence and parameter variations for the average photon energy in 
$B\to X_s\gamma$ decays. See text for explanation.}}}
\vspace{0.1cm}
\begin{center}
\begin{tabular}{|c|ccc|c|ccc|c|}
\hline\hline
$\langle E_\gamma\rangle$ [MeV] & $\mu_h$ & $\mu_i$ & $\mu_0$ & Comb.\
 & $m_b(\mu_*,\mu_*)$ & $-\lambda_1$ & $\alpha_s(M_Z)$ & Comb.\ \\
\hline\hline
2272 & $\pm 1$ & ${}_{\,-17}^{\,+19}$ & ${}_{\,-70}^{\,+48}$
 & ${}_{\,-72}^{\,+51}$ & $\pm 37$ & $\pm 10$ & ${}_{\,+6}^{\,-7}$
 & $\pm 39$ \\
\hline\hline
\end{tabular}
\end{center}
\end{table}

The last quantity we wish to explore is the average photon energy. As 
discussed in Section~\ref{sec:lowratios}, this quantity is almost insensitive 
to high-scale physics as well as to non-perturbative hadronic effects. 
However, it is very sensitive to the interplay of physics at the intermediate
and low scales, as illustrated by the approximate relation (\ref{Eavgapprox}). 
Our predictions for $\langle E_\gamma\rangle$ and its theoretical 
uncertainties are summarized in Table~\ref{tab:EavgScales} for the case 
$E_0=1.8$\,GeV, corresponding to the cut employed in \cite{Koppenburg:2004fz}. 
Since in this case the differences between the three variants of the 
shape-function scheme are insignificant, we only show results for RS~2. As 
expected, we find essentially no dependence on the hard matching scale, a 
modest dependence on the intermediate scale, and a more pronounced sensitivity 
to the low scale. The combined errors from scale variations are of order 
50--70\,MeV. The study of uncertainties due to parameter variations exhibits 
that the prime sensitivity is to the $b$-quark mass, which is expected, since 
$\langle E_\gamma\rangle=m_b/2+\dots$ to leading order. The next-important 
contribution to the error comes from the HQET parameter $\lambda_1$. The total 
error is about 40\,MeV.

Combining these results, we have to a very good approximation
\begin{equation}\label{Eavgres}
   \langle E_\gamma\rangle \Big|_{E_0=1.8\,{\rm GeV}}
   = (2.27_{\,-0.07}^{\,+0.05})\,\mbox{GeV}
   + \frac{\delta m_b}{2} - \frac{\delta\lambda_1}{4m_b} \,,
\end{equation}
where the error accounts for the perturbative uncertainty. The central values 
for the relevant input parameters are $m_b(\mu_*,\mu_*)=4.65$\,GeV and
$\lambda_1=-0.25$\,GeV$^2$, and the quantities $\delta m_b$ and 
$\delta\lambda_1$ parameterize possible deviations from these values. Our
prediction is in excellent agreement with the Belle result in 
(\ref{belle}). This finding provides support to the value of the $b$-quark 
mass in the shape-function scheme extracted in \cite{Bosch:2004th}. We stress, 
however, that the large perturbative uncertainties in the formula for 
$\langle E_\gamma\rangle$ impose significant limitations on the precision with 
which $m_b$ can be extracted from a measurement of the average photon energy.
Our estimate above implies a perturbative uncertainty of
$\delta m_b[{\rm pert.}]={}_{-100}^{+140}$\,MeV in the extracted value of 
$m_b$, which could only be reduced by means of higher-order calculations. This 
uncertainty is in addition to twice the experimental error in the measurement 
of $\langle E_\gamma\rangle$, which at present yields 
$\delta m_b[{\rm exp.}]=86$\,MeV.

\newpage
\section{Conclusions and outlook}

In this work, we have performed the first systematic analysis of the inclusive 
decays $B\to X_s\gamma$ in the presence of a photon-energy cut 
$E_\gamma\ge E_0$, where $E_0$ is such that $\Delta=m_b-2E_0$ can be 
considered large compared to $\Lambda_{\rm QCD}$, while still $\Delta\ll m_b$. 
This is the region of interest to experiments at the $B$ factories. The first 
condition ($\Delta\gg\Lambda_{\rm QCD}$) ensures that a theoretical treatment 
without shape functions can be applied. However, the second condition 
($\Delta\ll m_b$) means that this treatment is {\em not\/} a conventional 
heavy-quark expansion in powers of $\alpha_s(m_b)$ and 
$\Lambda_{\rm QCD}/m_b$. Instead, we have shown that three distinct 
short-distance scales are relevant in this region. They are the hard scale 
$m_b$, the hard-collinear scale $\sqrt{m_b\Delta}$, and the low scale 
$\Delta$. To separate the contributions associated with these scales requires 
a multi-scale operator product expansion (MSOPE), we which have constructed in 
this work. 

Our approach allows us to study analytically the transition from the 
shape-function region, where $\Delta\sim\Lambda_{\rm QCD}$, into the MSOPE
region, where $\Lambda_{\rm QCD}\ll\Delta\ll m_b$, into the region 
$\Delta={\cal O}(m_b)$, where a conventional heavy-quark expansion applies. 
This is a significant improvement over previous work. For instance, it has 
sometimes been argued that exactly where the transition to a conventional
heavy-quark expansion occurs is an empirical question, which cannot be 
answered theoretically. Our formalism provides a precise, quantitative answer 
to this question. In particular, for $B\to X_s\gamma$ with realistic cuts on 
the photon energy one is {\em not}\/ in a region where a simple short-distance 
expansion at the scale $m_b$ can be justified. The precision that can be 
achieved in the prediction of the $B\to X_s\gamma$ branching ratio is, 
ultimately, determined by how well perturbative and non-perturbative 
corrections can be controlled at the lowest relevant scale $\Delta$, which in 
practice is of order 1\,GeV. Consequently, we find larger theoretical 
uncertainties than previous authors. These uncertainties are dominated by yet 
unknown higher-order perturbative effects. Non-perturbative, hadronic effects 
at the scale $\Delta$ appear to be small and under control. 

Our treatment of the $B\to X_s\gamma$ branching ratio includes a complete 
resummation of logarithms $\ln(\Delta/m_b)$ at next-to-next-to-leading order 
in  renormalization-group improved perturbation theory. This level of 
precision has not been achieved before. Besides the calculations performed 
here and in \cite{Bosch:2004th,Lange:2003ff}, we have used multi-loop 
calculations for the cusp anomalous dimension 
\cite{Korchemsky:wg,Moch:2004pa}, the anomalous dimension of the shape 
function \cite{Korchemsky:1992xv} (which we have corrected, see Appendix~A.1
and also \cite{Einanprivate}), and the anomalous dimension of the 
leading-order current operator in soft-collinear effective theory 
\cite{inprep}. These ingredients are needed in order to achieve a complete 
separation of the perturbative corrections controlled by the three couplings 
$\alpha_s(m_b)$, $\alpha_s(\sqrt{m_b\Delta})$, and $\alpha_s(\Delta)$, which 
differ in magnitude by about a factor 2. Our prediction for the CP-averaged 
$B\to X_s\gamma$ with a cut $E_0=1.8$\,GeV is 
$\mbox{Br}(B\to X_s\gamma)=(3.38_{\,-0.42}^{\,+0.31}\,[\mbox{pert.}]\,%
{}_{\,-0.30}^{\,+0.32}\,[\mbox{pars.}])\times 10^{-4}$. With this cut 
$(89_{\,-7}^{\,+6}\,\pm 1)\%$ of all events are contained. The theory 
uncertainty we estimate is significantly larger than that found by previous 
authors, and this fact has important implications for searches of New Physics 
in radiative $B$ decays. Quite generally, the constraints on model parameter 
space have to be relaxed significantly. We have illustrated this fact with the 
example of the type-II two-Higgs-doublet model, for which we find that the 
lower bound on the charged-Higgs mass is reduced to approximately 200\,GeV.

This is not the first time in the history of $B\to X_s\gamma$ calculations 
that issues of scale setting have changed the prediction and error estimate 
for the branching ratio. In \cite{Czarnecki:1998tn}, Czarnecki and Marciano 
have pointed out that the electromagnetic coupling $\alpha$ in the expression 
for the decay rate should be identified with the fine-structure constant 
(normalized at $q^2=0$), and not with $\alpha(m_b)$ renormalized at the scale 
of the heavy quark in the decay. This lowered the prediction for the branching 
ratio by about 5\%. More recently, Gambino and Misiak have argued that the 
charm-quark mass, which enters the next-to-leading order corrections to the 
$B\to X_s\gamma$ rate via penguin loops, should be identified with a running 
mass $m_c(\mu)$ with $\mu\sim m_b$ rather than with the pole mass 
\cite{Gambino:2001ew}. This observation increased the prediction for the 
branching ratio by about 8\%, and at the same time it increased the error 
estimate associated with the value of the ratio $m_c/m_b$, which before had 
been taken to be the (rather well known) ratio of the two pole masses. The 
point we emphasize in the present work, namely that some effects in 
$B\to X_s\gamma$ decays should be described by the couplings 
$\alpha_s(\sqrt{m_b\Delta})$ and $\alpha_s(\Delta)$ (and power corrections at 
the scale $\Delta$) rather than $\alpha_s(m_b)$, is of a similar nature. 
However, in our case the change in perspective about the theory of 
$B\to X_s\gamma$ decay is more profound, as it imposes limitations on the very 
validity of a short-distance treatment. If the short-distance expansion at the 
scale $\Delta$ fails, then the rate {\em cannot\/} be calculated without 
resource to non-perturbative shape functions, which would introduce an 
irreducible amount of model dependence. In practice, while 
$\Delta\approx 1.1$\,GeV (for $E_0\approx 1.8$\,GeV) is probably sufficiently 
large to trust a short-distance analysis, it would be unreasonable to expect 
that yet unknown higher-order effects should be less important than in the 
case of other low-scale applications of QCD, such as in hadronic $\tau$ 
decays. 

Given the prominent role of $B\to X_s\gamma$ decay in searching for physics 
beyond the Standard Model, it is of great importance to have a precise 
prediction for its rate in the Standard Model. The present work shows that the 
ongoing effort to calculate the dominant parts of the next-to-next-to-leading 
corrections in the conventional heavy-quark expansion is only part of what is 
needed to achieve this goal. Equally important will be to compute the dominant 
higher-order corrections of order $\alpha_s^2(\Delta)$ and 
$\alpha_s^2(\sqrt{m_b\Delta})$, and to perform a renormalization-group 
analysis of the leading kinematic power corrections of order $\Delta/m_b$. In
fact, our error analysis suggests that these effects are potentially more
important that the hard matching corrections at the scale $m_b$. Let us finish 
by mentioning two possible approaches for addressing the issue of higher-order
perturbative effects at the intermediate and low scales: First, it would be
interesting to calculate the terms of order $\beta_0\alpha_s^2$ at the scales 
$\mu_i\sim\sqrt{m_b\Delta}$ and $\mu_0\sim\Delta$. While this would fall short 
of a complete calculation of ${\cal O}(\alpha_s^2)$ corrections, the ``BLM 
terms'' associated with the $\beta$ function are often numerically dominant 
\cite{Brodsky:1982gc,Neubert:1994vb}. We stress that the known 
${\cal O}(\beta_0\alpha_s^2)$ terms computed in the conventional heavy-quark
expansion \cite{Ligeti:1999ea} are not sufficient for this purpose. Separate 
computations of ${\cal O}(\beta_0\alpha_s^2)$ terms at the scales 
$\mu_h\sim m_b$, $\mu_i\sim\sqrt{m_b\Delta}$, and $\mu_0\sim\Delta$ would be 
required to perform a meaningful BLM scale setting. This statement is 
explained in Appendix~A.4. Secondly, the convergence of the perturbative 
expansion at the low scale $\mu_0\sim\Delta$ may be improved by borrowing the 
idea of ``contour resummation'' developed in \cite{LeDiberder:1992te}. Since 
the shape-function integrals can be written as contour integrals in the 
complex plane along a circle with radius $\Delta$, in may be more appropriate 
to use a contour-weighted coupling constant rather than the naive coupling 
$\alpha_s(\Delta)$. Exploring the numerical impact of these two proposals is 
left for future work.

\vspace{0.5cm}\noindent
{\em Acknowledgments:\/}
I am grateful to Joan Soto and Domenec Espriu for their hospitality during a 
visit to the Departament d'Estructura i Constituents de la Mat\`eria at 
Universitat de Barcelona, Spain, where part of this work was performed. It is 
a pleasure to thank Bj\"orn Lange, Xavier Garcia i Tormo, and Ignazio Scimemi
for useful discussions. I am grateful to Gregory Korchemsky for bringing the 
paper \cite{Grozin:1994ni} to my attention, and for discussions concerning the 
two-loop anomalous dimension of the shape function. Finally, I am indebted to 
Einan Gardi for providing the perturbative expansion of his expression for the 
$B\to X_s\gamma$ rate obtained in \cite{Gardi:2004ia}, and for many 
illuminating discussions of factorization results in deep-inelastic 
scattering, which have been instrumental in finding the (hopefully correct) 
expressions for the two-loop anomalous dimensions presented in Appendix~A.1. 
This research was supported by the National Science Foundation under Grant 
PHY-0355005.

\newpage
\section*{Appendices}

\subsubsection*{A.1~~Anomalous dimensions and RG functions}

The exact solutions (\ref{RGEsols}) to the RG equations in (\ref{dgl}) can be 
evaluated perturbatively by expanding the anomalous dimensions and $\beta$ 
function,
\begin{eqnarray}
   \Gamma_{\rm cusp}(\alpha_s) &=& \Gamma_0\,\frac{\alpha_s}{4\pi}
    + \Gamma_1 \left( \frac{\alpha_s}{4\pi} \right)^2
    +  \Gamma_2 \left( \frac{\alpha_s}{4\pi} \right)^3 + \dots \,,
    \nonumber\\
   \beta(\alpha_s) &=& -2\alpha_s \left[ \beta_0\,\frac{\alpha_s}{4\pi}
    + \beta_1 \left( \frac{\alpha_s}{4\pi} \right)^2
    +  \beta_2 \left( \frac{\alpha_s}{4\pi} \right)^3 + \dots \right] \,,
\end{eqnarray}
and similarly for the remaining anomalous dimensions. We work consistently at 
next-to-leading order in RG-improved perturbation theory, keeping terms 
through order $\alpha_s$ in the final expressions for the Sudakov exponent $S$ 
and the functions $a_\Gamma$, $a_\gamma$, and $a_{\gamma'}$. For $a_\Gamma$ 
one obtains the standard expression
\begin{equation}\label{asol}
   a_\Gamma(\nu,\mu)
   = \frac{\Gamma_0}{2\beta_0} \left[ \ln\frac{\alpha_s(\mu)}{\alpha_s(\nu)}
   + \left( \frac{\Gamma_1}{\Gamma_0} - \frac{\beta_1}{\beta_0} \right)
   \frac{\alpha_s(\mu) - \alpha_s(\nu)}{4\pi} + \dots \right] . 
\end{equation}
The result for the Sudakov factor $S$ is more complicated, as it is necessary 
to include terms of next-to-next-to-leading logarithmic order. We obtain
\begin{eqnarray}
   S(\nu,\mu) &=& \frac{\Gamma_0}{4\beta_0^2}\,\Bigg\{
    \frac{4\pi}{\alpha_s(\nu)} \left( 1 - \frac{1}{r} - \ln r \right)
    + \left( \frac{\Gamma_1}{\Gamma_0} - \frac{\beta_1}{\beta_0}
    \right) (1-r+\ln r) + \frac{\beta_1}{2\beta_0} \ln^2 r \nonumber\\
   &&\mbox{}+ \frac{\alpha_s(\nu)}{4\pi} \Bigg[ 
    \left( \frac{\beta_1\Gamma_1}{\beta_0\Gamma_0} - \frac{\beta_2}{\beta_0} 
    \right) (1-r+r\ln r)
    + \left( \frac{\beta_1^2}{\beta_0^2} - \frac{\beta_2}{\beta_0} \right)
    (1-r)\ln r \nonumber\\
   &&\hspace{1.35cm}
    \mbox{}- \left( \frac{\beta_1^2}{\beta_0^2} - \frac{\beta_2}{\beta_0}
    - \frac{\beta_1\Gamma_1}{\beta_0\Gamma_0} + \frac{\Gamma_2}{\Gamma_0}
    \right) \frac{(1-r)^2}{2} \Bigg] + \dots \Bigg\} \,,
\end{eqnarray}
where $r=\alpha_s(\mu)/\alpha_s(\nu)$. Whereas the two-loop anomalous 
dimensions and $\beta$ function are required in (\ref{asol}), the expression 
for $S$ also involves the three-loop coefficients $\Gamma_2$ and $\beta_2$.

\begin{figure}
\begin{center}
\epsfig{file=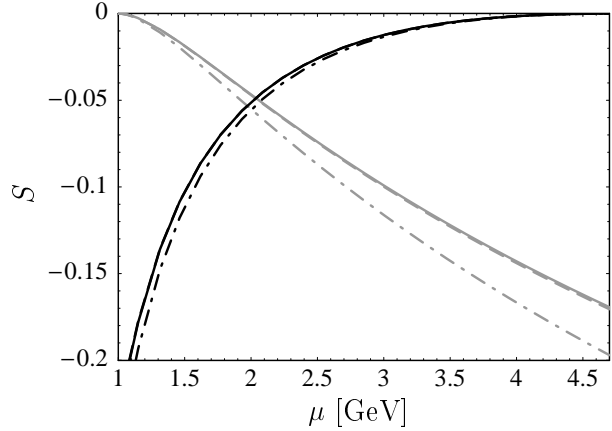,width=8cm}
\end{center}
\vspace{-0.4cm}
\centerline{\parbox{14cm}{\caption{\label{fig:Sudakov}
Sudakov exponents $S(m_b,\mu)$ (black) and $S(1\,\mbox{GeV},\mu)$ (gray) at 
next-to-next-to-leading order (solid), next-to-leading order (dashed), and 
leading order (dash-dotted). The solid and dashed curves are nearly 
indistinguishable.}}}
\end{figure}

The perturbative expansion of the QCD $\beta$ function to three-loop order is 
\cite{Tarasov:au} (all results refer to the $\overline{\rm MS}$ 
renormalization scheme)
\begin{eqnarray}
   \beta_0 &=& \frac{11}{3}\,C_A - \frac23\,n_f = \frac{25}{3} \,, \nonumber\\
   \beta_1 &=& \frac{34}{3}\,C_A^2 - \frac{10}{3}\,C_A\,n_f - 2 C_F\,n_f
    = \frac{154}{3} \,, \\
   \beta_2 &=& \frac{2857}{54}\,C_A^3 + \left( C_F^2 - \frac{205}{18}\,C_F C_A
    - \frac{1415}{54}\,C_A^2 \right) n_f
    + \left( \frac{11}{9}\,C_F + \frac{79}{54}\,C_A \right) n_f^2
    = \frac{21943}{54} \,, \nonumber
\end{eqnarray}
where the numerical values refer to $N_c=3$ and $n_f=4$. The two-loop 
coefficient of the cusp anomalous dimension has been known for a long time 
\cite{Korchemsky:wg}. However, its three-loop coefficient has only been 
calculated very recently by Moch et al.\ \cite{Moch:2004pa}. This is a lucky 
coincidence, because that calculation was done in a context not related to 
heavy-quark physics. The results are
\begin{eqnarray}
   \Gamma_0 &=& 4 C_F = \frac{16}{3} \,, \nonumber\\
   \Gamma_1 &=& 8 C_F \left[ \left( \frac{67}{18} - \frac{\pi^2}{6} \right)
    C_A - \frac{5}{9}\,n_f \right] 
    \approx 42.7695 \,, \nonumber\\
   \Gamma_2 &=& 16 C_F \Bigg[ \left( \frac{245}{24} - \frac{67\pi^2}{54}
    + \frac{11\pi^4}{180} + \frac{11}{6}\,\zeta_3 \right) C_A^2
    + \left( - \frac{209}{108} + \frac{5\pi^2}{27} - \frac{7}{3}\,\zeta_3
    \right) C_A\,n_f \nonumber\\
   &&\mbox{}+ \left( - \frac{55}{24} + 2\zeta_3 \right) C_F\,n_f
    - \frac{1}{27}\,n_f^2 \Bigg] \approx 429.507 \,.
\end{eqnarray}
Although the two- and three-loop coefficients of the $\beta$ function and cusp 
anomalous anomalous dimension are large, the perturbative expansion of the 
Sudakov exponent is extremely well behaved. This is illustrated in 
Figure~\ref{fig:Sudakov}, which shows the Sudakov exponents $S(\mu_h,\mu)$ and
$S(\mu_0,\mu)$ for $\mu_h=m_b=4.7$\,GeV and $\mu_0=1$\,GeV as a function of 
$\mu$.

The two-loop coefficient of the anomalous dimension $\gamma$ entering the 
shape-function evolution kernel in (\ref{gammaS}) has been calculated in 
\cite{Korchemsky:1992xv}. We have found some mistakes in the translation of 
the results for the two-loop graphs into the expression for the anomalous
dimension. The corrected result is \cite{privcom}
\begin{eqnarray}\label{littlegamma}
   \gamma_0 &=& - 2 C_F  = - \frac83 \,, \nonumber\\
   \gamma_1 &=& - 8 C_F \left[ \left( - \frac{37}{108} - \frac{\pi^2}{144}
    + \frac{9}{4}\,\zeta_3 - \frac{\kappa}{8} \right) C_A
    - \left( \frac{1}{54} + \frac{\pi^2}{72} \right) n_f \right] 
    \approx - 66.7531 + 4\kappa \,, 
\end{eqnarray}
where $\kappa=0$ under the assumption that the two-loop diagrams themselves 
were evaluated correctly in \cite{Korchemsky:1992xv}. However, there is reason
to believe that there might be an additional error in this paper, giving rise 
to a non-zero value $\kappa=4/3$ \cite{Einanprivate} (see also Appendix~A.2 
below), which we adopt in our numerical analysis.

The two-loop anomalous dimension for the leading-order SCET current operator
in (\ref{gamJ}) has not yet been computed directly. An analysis is in progress
and has already led to a prediction for the terms of order $C_F n_f$ 
\cite{inprep}. The remaining terms can be deduced by noting that the 
difference $\gamma^J\equiv\gamma'-\gamma$ is the non-cusp part of the 
anomalous dimension of the jet function \cite{Korchemsky:1994jb}, which is 
related to the familiar jet function from deep-inelastic scattering. We find
\begin{eqnarray}\label{gammaJ}
   \gamma_0^J &=& - 3 C_F \,, \nonumber\\
   \gamma_1^J &=& C_F \left[ \left( - \frac32 + 2\pi^2 - 24\zeta_3 \right) C_F
    + \left( - \frac{3155}{54} + \frac{22\pi^2}{9} + 40\zeta_3 \right) C_A
    + \left( \frac{247}{27} - \frac{4\pi^2}{9} \right) n_f \right] \nonumber\\
   &&\mbox{}+ (7-\pi^2) C_F\beta_0 \,,
\end{eqnarray}
where the terms in the first line in the expression for $\gamma_1^J$ are taken 
from \cite{Vogt:2000ci}, while the remainder in the second line is due to the 
non-trivial normalization of the SCET jet function in (\ref{Jres}). Combining 
the results (\ref{littlegamma}) and (\ref{gammaJ}), we obtain
\begin{eqnarray}\label{gammaprime}
   \gamma_0' &=& - 5 C_F = - \frac{20}{3} \,, \nonumber\\
   \gamma_1' &=& - 8 C_F \Bigg[ \left( \frac{3}{16} - \frac{\pi^2}{4}
    + 3\zeta_3 \right) C_F 
    + \left( \frac{1621}{432} + \frac{7\pi^2}{48} - \frac{11}{4}\,\zeta_3
    - \frac{\kappa}{8} \right) C_A \nonumber\\
   &&\mbox{}- \left( \frac{125}{216} + \frac{\pi^2}{24} \right) n_f
    \Bigg] \approx - 36.9764 + 4\kappa \,. 
\end{eqnarray}
Only the term proportional to $n_f$ in $\gamma_1'$ has so far been checked by 
a direct calculation in SCET \cite{inprep}.

\subsubsection*{A.2~~Perturbative expansion}

In this work, we have presented for the first time the complete RG-improved 
expression for the $B\to X_s\gamma$ decay rate, in which all logarithms 
$\ln\delta$ (with $\delta=\Delta/m_b$) are resummed at next-to-next-to-leading 
logarithmic order. In order to simplify the comparison of our result with 
those of other authors, we expand it in fixed-order perturbation theory and 
list the resulting terms at order $\alpha_s$ and $\alpha_s^2$. It suffices to 
focus on the perturbative correction to the term multiplying the product 
$m_b^3\,\overline{m}_b^2(\mu_h)\,|C_{7\gamma}(\mu_h)|^2$, where $m_b$ is the 
pole mass. We find
\begin{eqnarray}
   1 &+& C_F\,\frac{\alpha_s(m_b)}{4\pi}
    \left( 4\ln\frac{m_b}{\mu_h} - 2\ln^2\delta
    - 7\ln\delta - 5 - \frac{4\pi^2}{3} \right) \nonumber\\
   &+& C_F \left( \frac{\alpha_s(m_b)}{4\pi} \right)^2
    \left[ k_4 \ln^4\delta + k_3 \ln^3\delta
    + k_2 \ln^2\delta + k_1 \ln\delta + k_0 \right] + \dots \,,
\end{eqnarray}
where
\begin{eqnarray}
   k_4 &=& 2 C_F \,, \qquad 
   k_3 = 14 C_F + \frac{22}{3}\,C_A - \frac43\,n_f \,, \nonumber\\
   k_2 &=& \left( - 8\ln\frac{m_b}{\mu_h}
    + \frac{69}{2} + \frac{4\pi^2}{3} \right) C_F
    + \left( \frac{95}{18} + \frac{2\pi^2}{3} \right) C_A
    - \frac{13}{9}\,n_f \,, \nonumber\\
   k_1 &=&  \left( - 28\ln\frac{m_b}{\mu_h}
    + \frac{67}{2} + \frac{20\pi^2}{3} - 8\zeta_3 \right) C_F \nonumber\\
   &&\mbox{}+ \left( 2\kappa - \frac{953}{18} + \frac{34\pi^2}{9}
    + 4\zeta_3 \right) C_A
    + \left( \frac{85}{9} - \frac{4\pi^2}{9} \right) n_f \,.
\end{eqnarray}
These expressions are independent of the matching scales $\mu_i$ and $\mu_0$, 
and they have the correct dependence on $\mu_h$. The constant $k_0$ can only 
be obtained from a complete calculation of ${\cal O}(\alpha_s^2)$ corrections 
to the decay rate.

Resummed expressions for the $B\to X_s\gamma$ photon spectrum with 
next-to-leading logarithmic accuracy have been reported in 
\cite{Leibovich:1999xf} and \cite{Gardi:2004ia}. In the first paper, 
expressions for the coefficients $k_4$ and $k_3$ are derived, which agree with
our findings. In \cite{Gardi:2004ia}, Gardi has obtained a result for the
photon spectrum from which all four coefficients $k_i$ can be extracted. By 
matching his result for $k_1$ with ours, we conclude that $\kappa=4/3$.

\subsubsection*{A.3~~Kinematic power corrections for the average photon energy}

The functions $d_{ij}(\delta)$ entering the expression for the average photon 
energy in (\ref{Egamavg}) are given by
\begin{eqnarray}
   d_{77}(\delta) &=& \left( 8\delta^2 - \frac{14\delta^3}{3} + \delta^4
    \right) \ln\delta
    + \frac{7\delta^2}{2} - \frac{58\delta^3}{9} + 2\delta^4 \,, \nonumber\\
   d_{88}(\delta) &=& \frac49 \left[ \frac{\pi^2}{6} - L_2(1-\delta)
    + \left( \delta + \frac{\delta^2}{4} + \frac{\delta^3}{6} \right)
    \ln\delta - \delta - \frac{\delta^2}{4} - \frac{5\delta^3}{36}
    + \frac{\delta^4}{8} \right]  \nonumber\\
   &&\mbox{}+ \frac89 \left( \ln\frac{m_b}{m_s} - 1 \right)
    \left[ \ln(1-\delta) + \delta + \frac{\delta^2}{4} 
    + \frac{\delta^3}{6} \right] , \nonumber\\
   d_{78}(\delta) &=& \frac83 \left[ \frac{\pi^2}{6} - L_2(1-\delta)
    + \left( \delta + \frac{\delta^2}{2} \right) \ln\delta
    - \delta - \frac{7\delta^2}{8} + \frac{\delta^3}{6}
    - \frac{\delta^4}{16} \right] , \nonumber\\
   d_{11}(\delta) &=& - \frac89 \int_0^1\!dx\,
    (1-x)(1-x_\delta)^2 \left|\, \frac{z}{x}\,G\!\left(\frac{x}{z}\right)
    + \frac 12 \,\right|^2 , \nonumber\\
   d_{17}(\delta) &=& -3 d_{18}(\delta)
    = \frac43 \int_0^1\!dx\,x(1-x_\delta)^2\,
    \mbox{Re}\!\left[\, \frac{z}{x}\,G\!\left(\frac{x}{z}\right)
    + \frac 12 \,\right] ,
\end{eqnarray}
where $x_\delta=\mbox{max}(x,1-\delta)$, $z=(m_c/m_b)^2$, and the function 
$G(t)$ has been defined in (\ref{Gdef}).

\subsubsection*{A.4~~Comment on BLM scale setting}

Here we illustrate the simple fact that the BLM scale-setting procedure 
\cite{Brodsky:1982gc} for multi-scale problems is more complicated than in the 
familiar case with a single scale. Let us, for simplicity, ignore RG 
resummation effects due to anomalous dimensions and consider a physical 
quantity $A$, whose perturbative expansion is given by the product of two 
perturbative series at scales $M$ and $m$, with $M>m$. We write
\begin{equation}\label{Aorig}
   A = \left[ 1 + c_1\,a(M) + (2\beta_0 c_2 + c_2')\,a^2(M) + \dots \right]
   \left[ 1 + d_1\,a(m) + (2\beta_0 d_2 + d_2')\,a^2(m) + \dots \right] ,
\end{equation}
where $a\equiv\alpha_s/(4\pi)$, and $c_i^{(\prime)}$, $d_i^{(\prime)}$ are 
numerical coefficients. The BLM scales of the two series are
\begin{equation}\label{BLM1}
   \mu_{\rm BLM}^{\rm high} = M\,e^{-c_2/c_1} \,, \qquad
   \mu_{\rm BLM}^{\rm low} = m\,e^{-d_2/d_1} \,. 
\end{equation}
They are determined so as to absorb the ${\cal O}(\alpha_s^2)$ terms 
multiplying $\beta_0$ into the running coupling constants. Adopting the BLM 
philosophy, we would conclude that perturbation theory is well behaved as long 
as both $\mu_{\rm BLM}^{\rm high}$ and $\mu_{\rm BLM}^{\rm low}$ are in the 
perturbative regime. 

Imagine now that we compute $A$ in fixed-order perturbation theory using a 
single coupling constant $\alpha_s(\mu)$. We would obtain 
\begin{equation}\label{Aexp}
   A = 1 + (c_1 + d_1)\,a(\mu) + \left[ 2\beta_0 
   \left( c_2 + d_2 - c_1\ln\frac{M}{\mu} - d_1\ln\frac{m}{\mu} \right)
   + (c_2' + d_2' + c_1 d_1) \right] a^2(\mu) + \dots \,,
\end{equation}
and the associated BLM scale would be 
\begin{equation}\label{BLM2}
   \mu_{\rm BLM}^{\rm avg} = M \left( \frac{m}{M} \right)^{d_1/(c_1+d_1)}
   \exp\left( -\frac{c_2+d_2}{c_1+d_1} \right) . 
\end{equation}
Obviously, equation (\ref{Aexp}) does not provide the same information as 
(\ref{Aorig}), and in particular it does not allow us to compute the BLM
scales in (\ref{BLM1}). To this end we would need $c_2$ and $d_2$ separately, 
not just their sum. 

It is instructive to look at a couple of examples, where the conclusions 
derived from (\ref{Aexp}) would differ strongly from those derived from 
(\ref{Aorig}). Consider, for instance, a situation where $(c_1+d_1)$ is 
accidentally small. Then the BLM scale (\ref{BLM2}) is either very small or 
very large, whereas the BLM scales in (\ref{BLM1}) could be close to the 
scales $M$ and $m$, respectively. BLM scale setting based on the fixed-order 
calculation would then be totally misleading. Next, consider the case where 
the coefficient of $\beta_0$ in (\ref{Aexp}) is small, for instance by a 
particular choice of $\mu$. The fixed-order calculation would lead us to 
conclude that BLM-type corrections are small, whereas the BLM-type terms in 
(\ref{Aorig}) could still be large. Finally, consider the (somewhat 
pathological) example where $d_1\simeq c_1$ and $d_2\simeq -c_2$ with 
$\gamma=d_2/d_1>0$. Then the ``physical'' BLM scales are 
$\mu_{\rm BLM}^{\rm high}\simeq e^\gamma M$ and 
$\mu_{\rm BLM}^{\rm low}\simeq e^{-\gamma}\,m$. If $\gamma$ is large, 
perturbation theory may be in trouble, since $\mu_{\rm BLM}^{\rm low}$ may no 
longer be in the perturbative regime. Nevertheless, the ``average'' BLM scale 
$\mu_{\rm BLM}^{\rm avg}\simeq\sqrt{M m}$ is large, and the fixed-order 
calculation would thus indicate a well-behaved perturbative expansion.


\begin{thebibliography}{99}
\bibitem{Einanprivate}
E.~Gardi reports that he has confirmed (\ref{littlegamma}) with $\kappa=4/3$ 
from an independent calculation of the anomalous dimension $\gamma$ (private 
communication).
\end{thebibliography}
\end{document}